# Strategic Motivators for Ethical AI System Development: An Empirical and Holistic Model

Muhammad Azeem Akbar[1], Arif Ali Khan[2], Saima Rafi[3], Damian Kedziora[1, 4], Sami Hyrynsalmi[1]

[1]Software Engineering Department, Lappeenranta-Lahti University of Technology, 15210 Lappeenranta, Finland
[2]M3S Empirical Software Engineering Research Unit, University of Oulu, 90014 Oulu, Finland
[3]School of Computing, Engineering & the Built Environment, Edinburgh Napier University, EH10 5DT, Scotland
[4]Kozminski University, 03301 Warsaw, Poland

Corresponding author: azeem.akbar@lut.fi

## Abstract

**Context:** Artificial Intelligence (AI) offers significant opportunities to revolutionize industries and improve quality of life. However, ensuring its development is conducted responsibly is crucial to mitigating potential harms. Ethically robust AI systems require careful planning, governance, and the identification of key drivers that promote responsible AI development. **Objective:** The objective of this study is to identify and prioritize key motivators that drive the development of ethically responsible AI systems. **Method:** We conducted a Multivocal Literature Review (MLR) and a questionnaire-based survey to gather insights on prevalent practices in ethical AI development. Additionally, Interpretive Structure Modeling (ISM) was applied to examine relationships among the identified categories. The Matriced Impacts Croises Multiplication Appliquee Classement (MICMAC) analysis was utilized to classify these categories based on dependence and driving power. Finally, we employed the fuzzy Technique for Order Preference by Similarity to an Ideal Solution (TOPSIS) to prioritize key motivators. **Results:** Our study identified twenty key motivators categorized into eight related groups: *Human Resource, Knowledge Integration, Coordination, Project Administration, Standards, Technology Factor, Stakeholders, and Strategy & Matrices*. The ISM analysis revealed that *'Human Resource' and 'Coordination'* play key roles in influencing other categories. The MICMAC analysis further indicates that *CA1 (Human Resource), CA3 (Coordination), CA7 (Stakeholders), and CA8 (Strategy & Matrices)* belong to the independent cluster, suggesting these categories have strong driving power but weak dependence power. Interestingly, no category falls into the fully dependent cluster. Moreover, the results of fuzzy TOPSIS highlight critical actions such as *cultivating diverse teams, forming governance bodies for accountability, appointing ethical AI oversight leaders, and enforcing robust data collection and privacy standards* as the most critical motivators for developing ethically responsible AI systems. **Conclusion:** Organizations involved in AI development should consider the identified motivators to enhance policies and frameworks for responsible AI adoption.

## 1. INTRODUCTION

Artificial intelligence (AI) has become an essential phenomenon with significant impact on individuals, communities, and multiple industries, both white-collar and blue-collar [1]. The potential for AI to substantially transform society by offering technical and societal benefits, such as high productivity, efficiency, and enhanced user experience cannot be ignored. The ethical guidelines presented by the independent High-Level Expert Group On AI (AI HLEG, 2020) emphasize that AI should not be viewed as ultimate in itself but rather as a promising means to increase human flourishing, enhance individual and societal well-being, and promote progress and innovation [2]

Despite the possible (impressive) advantages of AI applications, concerns have been raised about their risks, as the complex and opaque systems may bring more social harm than benefits [1, 3]. As a result, there is a growing awareness of the need to consider the ethical implications of developing powerful and potentially life-altering AI systems. For instance, both the US government and numerous private enterprises got already cautious about utilizing AI-based decision-making software in such areas as healthcare, justice, employment, and creditworthiness, as they rely on virtual implications and might be intentionally or unintentionally coded with structural biases [1, 4].

As AI systems continue to advance, we have been witnessing also some ethical concerns. For example, Amazon's recruitment system was found to have a high rate of unsuccessful job applications that were biased against women applicants, which triggered discriminatory issues [5]. Since AI systems' decisions and recommendations can significantly impact people's lives, it is crucial to develop policies and principles that address their ethical aspects. Without such measures, the harms caused by AI-based software could jeopardize people's wellbeing, safety, control, and rights. Sustainable development of ethical AI systems requires not only technical efforts, but also consideration of the social, political, legal, and intellectual aspects. However, the current state of AI ethics is largely unknown to the public, practitioners, policymakers, and lawmakers [4, 6].

Along the rapid development of AI-based software, some ethical failures were also noticed. One example is the Amazon recruitment system, which rejected numerous job applications but was later discovered to be biased against women applicants due to its unequal selection criteria, resulting in discriminatory issues [3]. Given that AI systems' decisions and recommendations can significantly impact people's lives, it is crucial to develop policies and principles that address the ethical aspects of AI-based systems. Without proper ethical considerations, the harms caused by AI-based software could threaten people's control, safety, livelihoods, and rights. It is important to focus not only on the technical perspective of AI systems, but also consider their social, political, legal, and intellectual influence. However, the current state of AI ethics is not well understood by many public actors, as well as practitioners, communities, and policymakers [3, 7].

The AI system with ethical alignment should meet the following three requirements [8]: 1) compliance with the applicable laws and regulations, 2) adherence to ethical principles and values, and 3) technical and social robustness. There is a shortage of empirical studies of those requirements from the perspective of industrial practitioners and lawmakers. For instance, Vakkuri *et al.* [4] conducted a survey study to determine industrial perceptions based only on four AI ethics principles. Hunkenschroer and Christoph [9] interviewed researchers and practitioners to understand the implications of AI ethics principles and the motivation for enabling these principles in design practices. Similarly, Rakova *et al.* [10] focused on AI ethics guidelines, and recommend further conducting empirical studies on the subject. Also Halme et al. [11] highlighted the need to conduct more empirical studies on AI ethics, to further explore its principles and motivators (2024). Hence, our work strives to answer the following research questions:

**[RQ1]:** What motivators are important to consider for AI ethics reported in multivocal literature?
**[RQ2]:** What is the ranked-based holistic model of AI ethics motivators?

Additionally, the currently available guidelines and principles lack specificity to provide detailed information on particular areas to develop ethically robust AI systems. As a result, this research study aims to explore factors that could positively impact the ethical development of AI-based software and must be considered by practitioners to develop ethically sound systems. The study aims to serve as a body of knowledge for practitioners and research community by constructing new methods and techniques that are practically relevant for developing ethically sound AI-based systems.

## 2. BACKGROUND

AI has been discussed at a branch of applied ethics that deals with the ethical implications of developing and using information systems. It raises concerns about the impact of AI on human autonomy, freedom in democratic societies, and quality of life. Ethical considerations in the development of AI may facilitate the achievement of multiple societal goals. For example, they can promote the development of innovative technologies that foster ethical values and contribute to the collective well-being [12].

Developing ethically aligned and trustworthy AI-based systems can promote sustainable well-being in society by creating prosperity, maximizing wealth, and generating value [12]. By ensuring that AI software operates within acceptable ethical boundaries, we can establish a foundation of trust that encourages their widespread adoption. In this way, AI software can be harnessed to achieve positive social impact, in the same time mitigating any potential negative outcomes for society and individuals [13].

AI ethics can be understood as part of applied ethics, primarily focusing on the ethical aspects of software development and usage. It focuses on investigating how an AI-based system can impact human autonomy, freedom in a democratic society, and quality of life. The ethical reflection on AI technologies may serve multiple societal purposes [12]. It can shift the focus on innovations that foster ethical values and improve collective well-being. Ethically aligned and trustworthy AI can facilitate societal wellbeing by enhancing prosperity, maximization, and value creation [14]. However, the increasing popularity of AI systems has raised multiple concerns such as the reliability and neutrality of decision-making [14]. It must be ensured that AI technologies follow accountable decision-making process to execure scenarios ethically aligned with human values [15].

Various organizations and tech giants have formed committees to establish internal guidelines for AI ethics. For example, Google and SAP have introduced policies aimed at developing ethically aligned AI systems [16]. Likewise, the Association of Computing Machinery (ACM), Access Now, and Amnesty International have collaboratively proposed principles to foster ethically mature AI systems [16]. Additionally, the European Union's AI HLEG guidelines emphasize the importance of developing AI systems that are reliable and trustworthy [17]. Similarly, IEEE's Ethically Aligned Design (EAD) guidelines provide a framework of principles and recommendations that address both the ethical considerations and technical integrity of AI systems [18, 19]. Furthermore, the joint ISO/IEC international quality committee has introduced the ISO/IEC JTC 1/SC 42 standard, which encompasses the entire AI ecosystem, addressing trustworthiness, computational approaches, governance, standardization, and social concerns [20].

However, some researchers claim that proper guidelines and principles for AI ethics are not effectively adopted in across industries yet. McNamara *et al.*[20] conducted an empirical study to understand the influence of the ACM code of ethics on the software engineering decision-making process. Surprisingly, the findings reveal no evidence that it would regulate the decision-making activities at the studied sample. Vakkuri *et al.*[21] conducted multiple interviews to examine ethical practices in the AI industry. Their findings uncover that various guidelines are available, yet their deployment is far from being mature in the industry practice. Vakkuri et al. (2018) conducted a comprehensive overview of AI ethics principles and motivators [22], with an attempt to breach the gap between AI ethics research and practice.

## 3. RESEARCH METHODOLOGY

To formulate the aim of our study, we adopted a hybrid approach as presented in Fig 1. In the first step, we conducted a multivocal literature review (MLR) process to identify a comprehensive list of crucial motivators for ethical AI-based systems. In the second step, we validated the MLR findings by engaging industry experts through a questionnaire survey. To determine the relative significance of the identified motivators and their criticality for ethical AI systems, we used the ISM and Fuzzy TOPSIS approaches. We provide detailed descriptions of these methodological steps in the subsequent sections.

### A. Multivocal literature review (MLR)

We utilized the Multivocal Literature Review (MLR) [23] approach to address our research questions, which is a type of literature review that offers a comprehensive understanding from both the state-of-the-art and state-of-the-practice [23-25]. As illustrated at Figure 1, our research process involved scrutinizing literature from formal sources such as peer-reviewed journals and conferences, as well as informal sources such as industry standards, white papers, videos, blogs.

#### 1) Planning the review

In line with the MLR *approach* [23], we established a formal protocol before commencing the review process, and all research team members were involved in its different phases. MLR involves gathering data from two types of sources- academic, peer-reviewed literature, as well as grey literature, i.e. a type of literature that does not follow the quality control mechanisms before publication (Kamei et al., 2021), such as commercial, informal sources.

#### 2) Executing the review

To extract and consider the literature for data extraction, the research team jointly decided to follow the protocols described at the following sections.

##### a. Search string

We applied the guidelines provided by Zhang and Babar [26] to develop the search strings by concatenating keywords and their alternatives, following the below steps:

- Derive the major terms from Population, Intervention, and Outcome.
- Find synonyms and similar spellings of the derived terms.
- Verify these terms in various academic databases and search engines.
- The Boolean operators such as AND, and OR were used. AND operator was used to connect major elements of the search string (if allowed), and OR operator was considered to connect synonyms and similar spellings (if allowed).

Based on the above approach, the detailed search string elements are listed in Table I.

##### b. Digital platforms applied

To acquire the pertinent and valid literature, we implemented the developed search string across academic digital repositories (peer-reviewed literature), and search engines (grey literature). The platforms used for both forms of data are listed in Fig 1.

Table I: Search Terms

| Particular | Search Terms |
| --- | --- |
| SI1(Factors) | "Success factors" OR "factors" OR "aspects" OR "elements" OR "drivers" OR "motivators" OR "variables" |
| SI2 (Intervention) | *Artificial Intelligence:* ("artificial intelligence" OR "AI" OR "pattern recognition" OR algorithms) |
| | *Machine Learning:* ("machine learning" OR "predictive analytics" OR "pattern recognition" OR "deep learning." |
| SI3 (Intervention) | Ethics OR "human rights" OR "human values" OR "responsibility" OR "human control" OR "fairness" OR discrimination OR nondiscrimination OR "transparency" OR "explainability" OR "safety and security" OR "accountability" OR "privacy". |

| SI4 (Population) | "Software development community," "software operation community," "policymakers, lawyers," and "observers." |
|---|---|
| SI5 (Experimental) | *Formal Literature:* ("grounded theory," OR "interviews," OR "case studies," OR "questionnaire survey," OR "theoretical studies," OR "content analyses," OR "action research"). |
| | *Grey literature:* ("Videos," OR "Blogs," OR "white papers," OR "expert reports," OR "industry standards," OR "tweets," OR "website Q&A"). |
| "Final search string = (SI1) AND (SI2) AND (SI3) AND (SI4) AND (SI5)" | |

Additionally, we employed snowballing techniques (forward and backward) to track down the requisite data [27, 28]. In the context of our study, forward snowballing pertains to studies that cited the paper, whereas backward snowballing pertains to studies cited in the article's reference list [29, 30].

### c. *Inclusion criteria*

We employed the inclusion criteria derived from the guidelines outlined in [23, 31-33]. The essential components of our inclusion criteria are as follows: (i) the literature must be about AI ethics; (ii) the literature must address the influential factors of ethical AI, as well as social impact of AI, and AI principles; (iii) the literature must have an industry-oriented perspective; (iv) the literature must provide contextual information about the subject matter of the research; and (v) the literature must offer evident value to both academic and industrial researchers.

### d. *Exclusion criteria*

To eliminate irrelevant literature, we utilized exclusion criteria developed in line with the guidelines presented in [30, 33, 34]. The main components of our exclusion criteria are: (i) studies that lack relevance for researchers and industry professionals; (ii) literature not founded on primary evidence; and (iii) studies written in a language other than English.

### e. *Determination of Literature quality*

We evaluated the quality of the selected sources by adopting criteria from other Systematic Literature Review (SLR) studies in software engineering [23, 31, 32, 35, 36]. Our assessment followed the checklist of literature quality assessment criteria for both formal and gray literature, presented in Table II. Each chosen piece of literature was scrutinized against the assessment criteria using a Likert scale questionnaire, graded as "fully answered=1, partially answered=0.5, no answered=0". The final literature quality (LQ) score for each source of formal and grey literature is provided in Appendix-A.

### f. *Literature selection*

We followed a three-step process to select the most relevant literature based on the recommendations provided by Afzal et al. [30]. We started by using search strings to extract data from digital repositories. First the academic repositories, and then other digital repositories for grey literature. By executing the search string on digital libraries, we collected 4477 sources after employing inclusion and exclusion criteria. To further refine the literature source set, we applied the tollgate approach [30], and ultimately, 153 were deemed suitable for data extraction. Additionally, we identified 2 pieces of literature sources. The final set of selected literature is available at: https://tinyurl.com/7smz27vw

### g. *Data extraction*

To address our research questions, we applied the coding approach proposed by Kelle [37] to extract relevant data from the selected literature. We systematically labeled, grouped, and categorized the concepts, contributions, and findings from the collected data. This method proved particularly effective for analyzing qualitative datasets, enabling a thorough examination of the identified factors. After completing the data extraction process, we conducted an inter-rater reliability test to assess potential interpersonal bias. For this evaluation, three independent experts were involved in selecting 10 data sources (five academic publications and five from grey literature) and manually executing the entire data collection and extraction process. To measure inter-rater agreement between our data extraction team and the independent experts, we employed Kendall's nonparametric coefficient of concordance (W) [30]. The results indicated a strong level of agreement, with W = 0.91 (p = 0.002), as shown in Table III.

**h. *Data Synthesis***

For realizing our research objective, we identified a range of motivators positively impacting the development of AI systems that adhere to ethical principles. We also highlighted the essential process areas (or principles) that software development organizations should consider while designing AI-based software.

Table II: Quality assessment criteria

| Sr. No | Criteria for formal literature |
| --- | --- |
| Q1 | Does the selected work qualify as an empirical study? |
| Q2 | Are the study's aims and objectives clearly defined and logically justified? |
| Q3 | Is the research context described in sufficient detail? |
| Q4 | Does the research design effectively align with the study's objectives? |
| Q5 | Is the sample description adequate, including methods for selection and recruitment? |
| Q6 | Are the data collection methods suitable and well-documented? |
| Q7 | Are the data analysis techniques clearly explained and supported by the data? |
| Q8 | Has the researcher adequately considered their relationship with the participants? |
| Q9 | Are the findings clearly presented with credible results? |
| Q10 | Does the study offer valuable insights for research or practical applications? |
| **Sr. No** | **Criteria for grey literature** |
| Q1 | Is the publishing organization reputable? |
| Q2 | Is each author affiliated with a recognized institution? |
| Q3 | Does the author possess expertise in the subject area (e.g., principal software engineer)? |
| Q4 | Does the source clearly state its aim? |
| Q5 | Is a specific methodology outlined in the source? |
| Q6 | Are any limits clearly stated? |
| Q7 | Does the work focus on a particular population or case? |
| Q8 | Is the presentation of information balanced and unbiased? |
| Q9 | Are the statements in the source objective, or do they reflect subjective opinions? |
| Q10 | Is the publication date clearly stated? |

**3) *Reporting the review***

For the LQ assessment score, we evaluated the quality of the selected literature, to ensure the sources were suitable for addressing our research questions. Table II presents the LQ assessment logic for both types of literature. The determined QA (Quality Assessment) results indicated that over 78% of primary studies scored higher than 70%, and over 80% of grey literature scored at least 70%. These scores show that the selected literature effectively addressed our study's research questions. We used 50% as the threshold score. For more detailed information on the LQ assessment results, please see Appendix-A.

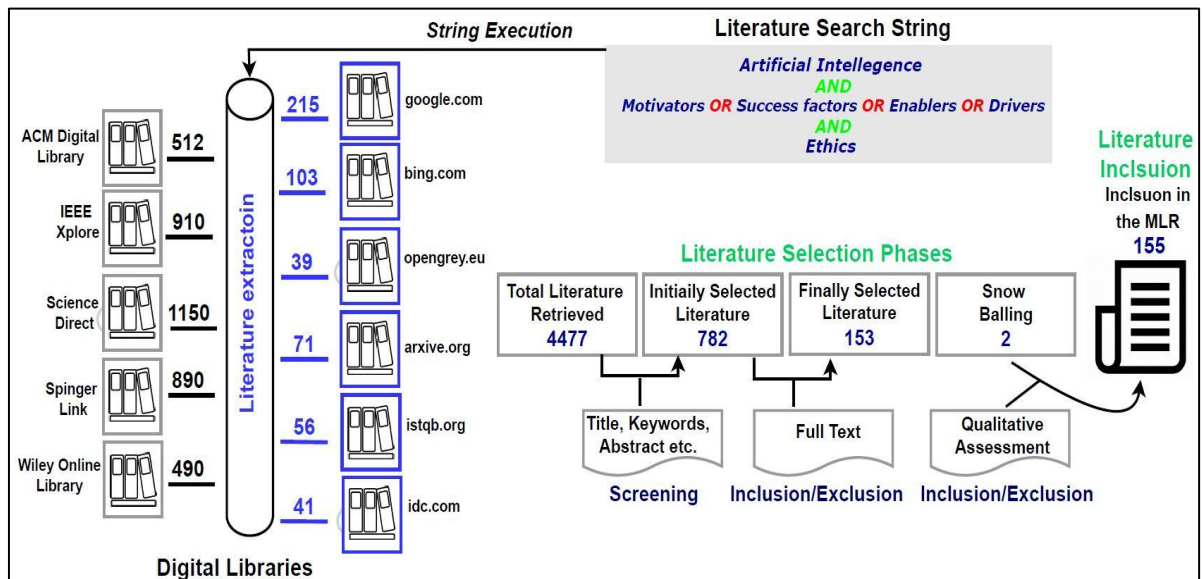

Fig 1: Selected literature for MLR

Table III: Kendall's Coefficient Of Concordance Test

| "Data Set" | "Kendall Chi-Squared" | "df" | "Subjects" | "Raters" | "*p*-value" | "W" |
|---|---|---|---|---|---|---|
| AI ethics | 35.434 | 14 | 10 | 3 | 0.001267 | 0.906761 |

```
"library(AIethics)
AIethics <- data.frame
(external_ex=c(3,3,3,4,3,4,2,3,3,2),
external_ex=(3,4,3,5,3,4,2,4,3,3),
external_ex3=(3,3,4,4,3,4,1,3,3,3)
authors_abc=(2,3,3,4,2,4,3,3,2,3)
)
KendallW(AIethics, TRUE)
KendallW(AIethics, TRUE, test=TRUE)
)
KendallW(t(d.att[, -1]), test = TRUE)
friedman.test(y=as.matrix(d.att[,-1]), groups = d.att$id)"
```

### B) Empirical research

Our empirical investigation aimed to analyze and characterize the relationship between the motivators identified in MLR (Machine Learning Research) and core categories of SPI (Software Process Improvement) manifesto [38] from the perspective of the AI-based software development community and policymakers. Our investigation aims to assist the software development community in creating ethically sound AI systems and contribute to the body of evidence in this area. Our research objective and RQ2 indicate that our investigation is exploratory [39]. For this purpose, we conducted an online questionnaire-based opinion survey to collect primary data. In the following subsections, we provide a detailed explanation of our survey's design and execution:

a) **Survey Development:** We designed an online survey based on the data collected from MLR (section 4.1), and we formulated a draft set of questions to include motivators for AI ethics adoption as defined in the SPI manifesto. The first draft of the questionnaire was developed based on the experience of our research team [40-42], and our consideration of questionnaire development guidelines, several survey guidelines, as well as experience reports [43-45]. The questionnaire draft was sent to five industry experts with an invitation for assessment and feedback. Our goal was to ensure that the terminology used in the questionnaire was understandable to industry professionals. Discussing and agreeing on consistent vocabulary is important when conducting joint projects or studies between academia and industry [40-42]. The early feedback from practitioners was used to refine the set of questions. After finalizing the survey, we had 34 questions that are provided in Appendix B. The first set of 12 questions covered the profiles and demographics of the participants, projects, and companies. Most of them had pre-designed single or multiple-response answers. Aside that, participants were able to provide open answers to several questions (e.g. the participant's current position) as they could have been more than one single role in their organization. The remaining questions focused on matters related to AI ethics adoption in the SPI manifesto context. We used the five-point Likert scale form (from strongly disagree to strongly agree), maintaining neutral perspective in order to collect unbiased data.

b) **Survey Execution:** For data collection, we utilized Google Forms (docs.google.com/forms) as the survey tool hosted on Google Drive (drive.google.com). Before starting the data collection process, we obtained an approval from LUT University's Research Ethics Board. The survey was made available to participants from February to April 2024, with voluntary and anonymous participation. Respondents had the option to withdraw from the survey at any time. To reach as many responders as possible, we developed a publicity and execution plan. We sent email invitations to our network of partners and contacts in 63 software companies around the world. Following the snowballing, we asked participants to forward the invitation email (with survey URL) to their contacts. We also made public invitations to the software engineering community through social media, such as LinkedIn, Facebook, and Twitter, and by sending emails to the management offices of over 16 Technology Development Zones, as well as Research Finland Agency. To avoid duplicate invitations and to ensure that we communicated our invitation with each organization through a single point of contact (SPOC), we established a publicity schedule for each target sector and contact person. Although we did not keep track of the response rates due to the anonymity of the survey, the iterative publicity schedule allowed us to monitor changes in the survey distribution and estimate response rates after each iteration.

It is worth noting that, as with other online software engineering surveys [40-42], some participants left some questions unanswered. However, even partial answers were relevant to this study, as they also provided valuable information. In total, we received 156 complete responses that were used for data analysis, and we are willing to make the data available upon request for replicability purposes, as well as further analysis by other researchers.

c) **Data analysis:** At this step, we only analyzed the 156 responses that were 'complete' (with all answers submitted by a responder). We applied the frequency analysis method to inspect the descriptive data types [46]. We comparatively analyzed the survey variables and computed the agreement level between the survey participants based on the selected Likert scale. Such approach has been applied to similar data analyses by other researchers, such as Niazi *et al* [46]., as well as Ali and Khan *et al.,* Akbar *et al.,* Keshta *et al.,* Mahmood *et al.* [31, 47-50].

### C)  Phase 2: ISM Approach

According to Sage [51], Interpretive Structural Modeling (ISM) is a methodology that helps to bring order and direction to the complex relationship between factors and systems, resulting in a holistic, systematic model. ISM is an interactive learning technique designed to structure factors that are directly or indirectly related within a holistic model, presenting a conceptual representation in a clear graphical format [51, 52]. This method effectively addresses the complexities associated with relationships among different factors, offering a better understanding of these interconnections. Numerous studies have applied ISM to develop conceptual models for analyzing factor relationships, including the works of Kannan et al. [53], Sharma et al. [54], and Agarwal and Vrat [55]. However, the results of ISM can be affected by the interpersonal bias in the experts' opinions. Additionally, ISM does not give weights to analyze the ranking of each factor at a level. To address these concerns, we then adopted the fuzzy TOPSIS (Technique for Order Preference by Similarity to Ideal Situation) approach, prioritizing the identified motivation factors based on their relationship with ten knowledge areas of AI ethics. In order to identify the interaction between core areas of AI ethics motivators, we followed the steps adopted in the ISM approach. Fig 1 presents the detailed steps we adopted to follow the ISM approach, which was based on the study of Raj and Attri [56].

### D)  Phase 3: Fuzzy TOPSIS

TOPSIS approach was developed in 1981 by Hwang and Yoon [57] with the aim of defining positive and negative ideal solutions. The positive ideal solution maximizes benefit criteria and minimizes cost criteria, whereas the negative ideal solution maximizes cost criteria and reduces benefit criteria [46]. A solution is considered as the best if it is close to the positive ideal solution and far from the negative ideal solution. To address multi-criteria decision-making by collecting experts' opinions on a specific subject, Chen and Tsao [58] developed fuzzy TOPSIS approach, where decision-makers assign weights to each criterion using linguistic variables, and then the responses are converted into triangular fuzzy numbers (TNFs). According to Rafi *et al.*, Kanna *et al.*, as well as Krohling and Campanharo [59-61], TNFs are useful for managing the vagueness of linguistic terms used by decision-makers. The fuzzy TOPSIS algorithms we deployed for addressing multi-criteria decision-making are provided below.

**Step 1.** Suppose the problem with m alternative and n criteria can be expressed in a matrix form, where A1, A2,….Am are the alternatives, E1, E2,…, En is the evaluation criteria, Fij is the performance rating judged by decision-makers to the alternative Fi against the criterion Ej, and Wj of each criterion Ej represents weight.

$$D = (F_{ij})_{m*n} = \begin{array}{c} A_1 \\ A_2 \\ \vdots \\ A_m \end{array} \begin{array}{cccc} E_1 & E_2 & \dots & E_n \\ \begin{pmatrix} A_{11} & A_{12} & \dots & A_{1n} \\ A_{21} & A_{22} & \dots & A_{2n} \\ \vdots & \vdots & \vdots & \vdots \\ A_{m1} & A_{m2} & \dots & A_{mn} \end{pmatrix} \end{array} \quad (1)$$

**Step 2.** Assemble the alternatives and their weighted criteria according to (1). Assign the ratings to each defined criterion and their alternatives using Bozbura *et al.* [62] fuzzy triangular scale, as shown in Table IV.

Table IV: Triangular-fuzzy scale *[62]*

| Linguistic terms | Triangular Fuzzy Scale |
| --- | --- |
| "Just Equal (JE)" | "(1,1,1)" |
| "Equally Important (EI)" | "(0.5,1,1.5)" |
| "Weakly Important (WI)" | "(1,1.5,2)" |
| "Strongly More Important (SMI)" | "(1.5,2,2.5)" |
| "Very Strongly More Important (VSMI)" | "(2,2.5,3)" |
| "Absolutely more important (AMI)." | "(2.5,3,3.5)" |

**Step 3.** Determine the aggregate fuzzy rating of K decision-makers for each criterion by using equations (2) and (3).

$$A_{ij} = K^{min}\{x_{ijk}\}, b = \frac{1}{K}\sum_{k=1}^{K} y_{ijK}, c = K^{max}\{z_{ijK}\} \qquad (2)$$

$$W_{j1} = K^{min}\{x_{jK1}\}, b = \frac{1}{K}\sum_{k=1}^{K} y_{jK2}, c = K^{max}\{z_{jK3}\} \qquad (3)$$

Where $A_{ij} = (x_{ij}, y_{ij}, z_{ij})$ and i= 1,2,3,…m and j=1,2,3…,n and weight of each criterion is calculated as

$$W_j = (W_{j1}, W_{j2}, W_{j3}).$$

**Step 4.** Calculate the normalized decision matrix of 'R' using linear scale transformation (4). The matrix after normalization will present as follows:

$$R\~= [r_{ij}]_{m*n} \qquad (4)$$

Equation (5) and (6), given below, are used to calculate each alternative's cost and benefit criteria.

$$r_{ij} = \left(\frac{x_{ij}}{z_j^+}, \frac{y_{ij}}{z_j^+}, \frac{z_{ij}}{z_j^+}\right) \text{ and } z_j^+ = \max_i z_{ij} \text{ (benefit criteria)} \qquad (5)$$

$$r_{ij} = \left(\frac{x_j^-}{z_{ij}}, \frac{x_j^-}{y_{ij}}, \frac{x_j^-}{x_{ij}}\right) \text{ and } x_j^- = \max_i l_{ij} \text{ (cost criteria)} \qquad (6)$$

**Step 5.** Compute the weighted normalized fuzzy decision matrix V~, by multiplying the weights of each criterion Wj with calculated values of the normalized fuzzy decision matrix (7).

$$V \cong [v_{ij}]_{m*n}, \qquad (7)$$

where $v_{ij}$ is calculated by using (8)

$$v_{ij} = A_{ij} * W_j \qquad (8)$$

**Step 6.** This step calculates the fuzzy positive and negative ideal solutions as shown in (9) and (10).

$$A^+ = [v_{1,}^+ v_{j,}^+ …..v_m^+] \qquad (9)$$

$$A^- = [v_{1,}^- v_{j,}^- …..v_m^-] \qquad (10)$$

The values of positive and negative ideal solutions range between [0.1].

**Step 7.** Compute the distance of each alternative from a positive and negative ideal solution using (11) and (12).

$$D_i^+ = \sum_{j=1}^n D_v(v_{ij}, v_j^+) \qquad (11)$$

$$D_i^- = \sum_{j=1}^n D_v(v_{ij}, v_j^-) \qquad (12)$$

Where D represents the distance between two fuzzy numbers.

**Step 8.** To compute the ranks of alternatives CCi value is calculated by considering a fuzzy positive and negative ideal solution using (13).

$$CCi = \frac{D_i^-}{D_i^+ + D_i^-} \qquad (13)$$

**Step 9.** Define the priority-wise ranking of all alternatives according to CCi value. The higher the CCi value, the greater is the rank of that alternative.

## 4. RESULTS AND DISCUSSION

This section presents the discussion and analysis of the results. The identified motivators are detailed in Section 4-A, while Section 4-B explores the results of the ISM approach. Section 4-C discusses the findings from the Fuzzy TOPSIS analysis, and the proposed holistic model of AI ethics motivators is introduced in Section 4-D.

### A. List of Motivators for Development Organizations

By performing the multivocal literature review, we identified the motivators important for developing ethically robust AI systems, as depicted in Table V.

*Assign a leader responsible and accountable for ethical AI (M1)*
The importance of ethical facets in AI systems implies the need for building relevant leadership capabilities, such as Chief Ethics Officer, who ensure ethical issues are prioritized at all organizational actions. However, according to a recent report by Kuziemski and Misuraca, only 54% of organizations define such a role [1]. Its lack can lead to the negligence of ethical AI issues, and to prevent that, it is necessary to align business and technology leaders to clearly define responsibility for the ethical aspects of AI applications and products.

*Build diverse teams (M2)*
Previous studies pointed out a relationship between the lack of diversity in AI based system development team, as well as discriminatory behavior of some AI systems [1]. This reflects the need to build diverse and inclusive teams to design, develop, test, and maintain AI-based systems [19]. These teams should include individuals who vary in race, educational background, gender, and perspectives [1, 19]. Different viewpoints of the involved teams may contribute to embedding sensitivity in AI-based systems' design [1].

*Education and awareness to foster an ethical mind-set (M3)*
Ethics in AI cannot be secured sufficiently without establishing an ethical organizational culture [20]. This requires all the organizational entities to be aware of AI ethical issues and ethical framework to raise ethical concerns at every stage of the AI system's development. At a larger scale, knowledge regarding the potential ethical impacts of AI systems should be fostered across society [3]. Hence, education and training are important to raise public awareness regarding ethical issues and increase civic participation in the design of AI systems. However, to successfully educate, it is necessary to professionalize ethicists in the AI domain [3].

*Ethics council development (M4)*
A governance board has been recommended to oversee the organizational AI strategy and define the AI ethical framework [9, 20]. Such board should include external subject matter experts (SMEs) and ethicists. It should perform the following tasks [20]: (i) monitor how employees deal with ethical issues, (ii) address legal and regulatory risks, (iii) align the strategy of AI ethics to specific systems and products in the organization.

*Surveillance practices for data gathering and privacy of court users (M5)*
Commercial and governmental data gathering, and surveillance are treated as separate rather than intrinsically and inextricably linked. This miscasting has critical implications. When the debate considering such misunderstandings is recast, the bar for the justification of surveillance is raised. A new balance needs to be found in political, legal, and decision-making debate. The dispute over the surveillance's nature of the internet has often been considered misleading [22]. Privacy has been considered an individual right, as opposed to the need for security which is collective. In addition, the impact of data gathering, and surveillance has been limited to privacy without considering other human rights, such as freedom of expression. This impact occurs only when human examinations are performed without considering algorithmic examinations. Finally, the interrelations between commercial and governmental data gathering and surveillance have not been discussed. Instead, they are considered separate processes, while they should be discussed in political debates and decision-making contexts [22].

*Requirements of trustworthy AI (M6)*
A set of concrete requirements should be established to enable trustworthy AI. Such requirements are relevant to various stakeholders of the AI systems' life cycle, such as developers, implementers, end-users, and society at a larger scope [3]. For instance, developers are responsible for applying the requirements in the design and development processes. Implementers, i.e., public or private organizations that adopt AI-based systems within their business processes, should ensure that the systems meet the requirements of trustworthiness. Finally, the requirements should be communicated to end-users and society more broadly.

*Regulatory compliance and standardization (M7)* Organizations should set up a department responsible for ensuring AI systems comply with existing legal and regulatory frameworks [3]. In addition, such department should be responsible for continuous updating of AI systems' policies following technological and regulatory changes. Those include ISO and IEEE P7000 standards, whose future versions may also apply to AI systems' safety, technical robustness, and transparency [3]. Furthermore, users of AI systems can enhance their ability to recognize ethical conduct by using design and manufacturing standards or business practices, such as quality management systems [3]. For instance, employees should comply with the organizational code of conduct and business ethics which aim to encourage honest, legal, and ethical practices [21]. Ethical concerns and complaints must be submitted confidentially by employees and investigated by dedicated audit committee [21].

*Algorithmic accountability (M8)*
Despite the high potential of AI to globalize domains, such as healthcare [58], the trustworthiness and effectiveness of AI-based systems depend on the datasets used to train the algorithms. The more complete, updated, and representative the data sets are, the lower is the probability of algorithmic bias and discrimination. Such biases can be detected and mitigated using some online tools. Moreover, it is important to ensure robust data and privacy protection, as well as transparency to users by ensuring they understand how their data is being used [9]. The risks of biases should be evaluated at every stage of the product development cycle, particularly when choosing the technologies and datasets to train the algorithms in terms of their quality and diversity [23].

*Develop ethics and privacy roadmap (M9)*
AI systems may raise privacy concerns as they collect vast user data [2]. For example, it may enable AI systems to deduce the preferences of individuals based on their profiling data, such as age, gender, sexual orientation, and political views. Individuals can trust the data collection process only if they are ensured that this data will not be used against them in an unlawful or unfair manner [3]. In this regard, regulators are putting particular emphasis on data security standards. From an organizational perspective, it is necessary to establish appropriate data governance roadmap that consider data quality, integrity, domain relevance, access protocols, and processing mechanisms, whereas protecting privacy [3].

*Establish AI ethics principles tailored to the business domain (M10)*
AI ethics in organizations should be approached by establishing its high-level principles [9]. These principles should be tailored to the business, industry domain, and specific technology the organization uses. Their ultimate goal is to align AI interactions and organizational values. To achieve this goal, principles should be communicated at every business level through training, reward systems, and intensive leadership communications [9]. This approach may empower employees and encourage them to raise ethical concerns whenever they arise.

*AI ethics risk assessment policies (M11)*
Organizations should develop AI ethics risk assessment policies to predict and prevent ethical issues [9]. This may require incorporating changes into operational frameworks and developing an AI governance model that encompasses short, medium, and long-term goals and guidance regarding AI. This plan should be managed by a dedicated, qualified person responsible for the regular updating of the plan.

*Consider ethics by design (M12)*
Whereas it is true that ethical concerns related to technologies cannot be eliminated, it is also true that these concerns can be minimized by applying the 'ethics by design' principle [24]. This principle entails considering ethical issues, besides security and privacy, early in the design process. The early identification and mitigation of such issues shall result in long run benefits, such as time and cost savings.

*Automating ethics (M13)*
Increasing the decision-making power of AI agents, commonly referred to as 'techno-empowerment', implies the need to embed ethical standards into their performance [7]. Automating elements of AI ethics into their design can enable more detailed analysis and decision-making process. Nevertheless, it may face some resistance from humans, as AI will operate faster than humans, and consequently, humans will not be able to control the process.

*Erosion of privacy (M14)*
Tech companies collect a vast amount of private data from customers when they use digital products or services [25]. The initial goal of collecting this data was to train AI algorithms for targeted and customized advertising. However, ethical issues arise when the data is used for other purposes, such as the creation of employment offers without users' consent. Recent studies revealed that 60% of customers raised concerns regarding the use of their personal information. To establish customer trust, companies should ensure transparency about using the collected information, build appropriate mechanisms for consent, and protect customers' rights [25].

*Set up a governance body to implement measures of accountability (M15)*
Given the complexity of AI systems, it is of utmost importance to establish governance measures to leverage the benefits of AI systems and mitigate their negative consequences. It can be established by a governance board accountable for monitoring the AI strategy and defining the AI ethical framework [9].

*Screen the data used to train the AI system for bias (M16)*
The input data has a significant impact on AI systems. Biased input data can lead to wrong decisions and codifying racial or gender prejudices into the system. Therefore, it is important to ensure that the data used to develop AI systems is unbiased and data training is not affected by the implicit preferences of AI designers [63].

*Controllable AI with clear accountability (M17)*
Auditability and accountability are necessary facets of AI systems. Organizations must set up an accountability structure that defines the ownership and controls for complex emergency response. For instance, in case of an illegal trade execution by an AI agent, a company may lose millions of dollars, and it requires a clearly defined entity to take accountability and respond to it. In addition, an AI system should be auditable regarding transparency and explainability [9].

*Optimize guidance and tools (M18)*
Explainability of AI systems is required particularly when the decisions made by those systems have strong impact on someone's life and wellbeing [20]. However, the transparency decreases with an increase in prediction accuracy [64]. In this context, product managers need to decide how to make tradeoffs. Customized tools may facilitate the decision-making process by assessing the importance of explainability and accuracy for the specific system and providing product managers with suggestions on which features should be implemented in that system.

*Two-way communication with users (M19)*
Ethics are principles that guide behavior and interpretations [26]. Given the limited visibility of algorithms and the fact that humanoid robotic devices impact users' responses in unintelligible ways, it is important to adopt a user-participatory design. Participatory design involves two-way communication with users to design AI systems ethically [65].

*AI collides with patent law (M20)*
The influence of AI on copyright law has been discussed in various settings. For instance, following the 'selfie-taking monkey case', the United States Copyright Office modified the interpretation of "authorship" to outline that it would not register machine-produced works or any artifacts generated from a random or automatic mechanical process. However, given AI's technological and societal implications, additional discussions on the topic are paramount to facilitate necessary modifications of the patent systems. Such modifications should allow the system to achieve its objectives whereas preventing negative ethical, economic, and social effects [66].

Table V: Motivators for development organizations

| Core Areas | Motivators for AI-based software development practitioners |
|---|---|
| Human resource (CA1) | Assign a leader responsible and accountable for ethical AI (M1) |
| | Build diverse teams (M2) |
| Knowledge integration (CA2) | Education and awareness to foster an ethical mind-set (M3) |
| | Consider ethics by design (M12) |
| Coordination (CA3) | Erosion of privacy (M14) |
| | Two-way communication with users (M19) |
| | Surveillance practices for data gathering and privacy of court users (M5) |
| Project administration (CA4) | Establish AI ethics scope according to the business domain (M10) |
| | Regulatory compliance and standardization (M7) |
| | Requirements of trustworthy AI (M6) |
| Standards (CA5) | Develop ethics and privacy roadmap (M9) |
| | Controllable AI with clear accountability (M17) |
| | Algorithmic accountability (M8) |
| Technology factor (CA6) | Automating ethics (M13) |
| | Optimize guidance and tools (M18) |
| Stakeholders (CA7) | Ethics council development (M4) |
| | Set up a governance body to implement measures of accountability (M15) |
| | AI collides with patent law (M20) |
| Strategy and matrices (CA8) | Screen the data used to train the AI system for bias (M16) |
| | AI ethics risk assessment policies (M11) |

## B. Results of ISM approach

The Interpretive Structural Modeling approach we applied allowed us to examine interactions between the key areas of AI ethics motivators. Various researchers have applied ISM approach to study contextual interaction of some elements [53-55]. In order to develop the contextual interaction between the criteria, creating a structural-self-interaction matrix (SSIM) is essential, as presented in the following sections.

### 1) Structural-self-interaction matrix (SSIM)

Using expert opinions, we employed the ISM approach to investigate contextual relationship between the key categories of AI Ethics motivators. To collect these opinions, we formed an expert group and sent invitation letters to participants from the first survey. Ten experts with industry and R&D backgrounds in AI ethics agreed to participate in the decision-making process. Appendix-C provides a brief demographic information about the participants. While the sample size may limit the possibility of generalizing our study, we found that previous studies, such as Kannan *et al*. [53], Soni *et al*. [67], and Attri *et al*. [68], also used a small number of experts to develop their models. Having considered those experts' opinions, we developed the SSIM matrix, which uses the following symbols to indicate the direction of a relationship between AI ethics motivators (m and n):

a) "V" indicates a relationship from m motivator to n motivator.
b) "A" indicates a relationship from n motivator to m motivator.
c) "X" denotes a situation where both enablers m and n reach each other.
d) "O" signifies a situation with no relationship between motivator m and motivator n.

Table VI presents the SSIM developed based on the experts' opinions.

TABLE VI
SSIM MATRIX

| | CA8 | CA7 | CA6 | CA5 | CA4 | CA3 | CA2 | CA1 |
|---|---|---|---|---|---|---|---|---|
| CA1 | V | V | O | V | V | X | O | * |
| CA2 | V | O | O | O | O | O | * | * |
| CA3 | V | O | O | O | O | * | * | * |
| CA4 | V | V | X | X | * | * | * | * |
| CA5 | V | O | V | * | * | * | * | * |
| CA6 | O | O | * | * | * | * | * | * |
| CA7 | X | * | * | * | * | * | * | * |
| CA8 | * | * | * | * | * | * | * | * |

Based on the findings outlined in Table VI, a positive relationship appears between CA1 (Human resource) and CA8 (Strategy and matrices), as it is depicted by a 'V.' This suggests that a close working relationship between these two areas is necessary to facilitate improvements in AI-based systems. Similar relationship has been observed between CA1 (Human resources) and CA7 (Stakeholders). However, experts indicated no relationship between CA1 (Human resource) and CA6 (Technology factor), as it is represented by 'O'. This demonstrates that a lack of collaboration between those areas does not significantly impact the consideration of ethics in AI-based systems. Additionally, CA1 (Human resource) and CA3 (Coordination) were found to have a two-way relationship, which suggests that both of these areas contribute to the overall improvement of AI ethics. Interestingly, there was no 'A'-type relationships among the key areas of AI ethics motivators.

### 2) Reachability matrix

To create the reachability matrix, we converted the values of V, A, X, and O in the SSIM matrix into binary digits (0, 1). The following protocols were used to develop the reachability matrix:

a) If the value of m and n in SSIM is V, we replaced it with 1; otherwise, we assigned a value of 0.
b) If the value of m and n in SSIM is A, we replaced it with 0; otherwise, we assigned a value of 1.
c) If the value of m and n in SSIM is X, we replaced it with 1, and assigned a value of 1 to both the n and m entries.
d) If the value of m and n in SSIM is O, we replaced it with 0, and assigned a value of 0 to both the n and m entries.

By applying the above protocols, we developed the reachability matrix presented in Table VII.

Table VII: Reachability matrix

|      | CA1 | CA2 | CA3 | CA4 | CA5 | CA6 | CA7 | CA8 |
|------|-----|-----|-----|-----|-----|-----|-----|-----|
| CA1  | 1   | 0   | 1   | 1   | 1   | 0   | 1   | 1   |
| CA2  | 0   | 1   | 0   | 0   | 0   | 0   | 0   | 1   |
| CA3  | 1   | 0   | 1   | 0   | 0   | 0   | 0   | 1   |
| CA4  | 0   | 0   | 0   | 1   | 1   | 1   | 1   | 1   |
| CA5  | 0   | 0   | 0   | 1   | 1   | 1   | 0   | 1   |
| CA6  | 0   | 0   | 0   | 1   | 0   | 1   | 0   | 0   |
| CA7  | 0   | 0   | 0   | 0   | 0   | 0   | 1   | 1   |
| CA8  | 0   | 0   | 0   | 0   | 0   | 0   | 1   | 1   |

We determined the final reachability matrix to incorporate transitivity, as discussed in Section 3.3. The value of 1* was used to fill gaps in the data collected from experts whilst developing the SSIM matrix. The incorporation of the transitivity check is presented in Table VIII.

Table VIII: Transitivity check

|      | CA1 | CA2 | CA3 | CA4 | CA5 | CA6 | CA7 | CA8 | Div | rank |
|------|-----|-----|-----|-----|-----|-----|-----|-----|-----|------|
| CA1  | 1   | 0   | 1   | 1   | 1   | 1*  | 1   | 1   | 7   | 5    |
| CA2  | 0   | 1   | 0   | 0   | 0   | 0   | 1*  | 1   | 3   | 2    |
| CA3  | 1   | 0   | 1   | 1*  | 1*  | 0   | 1*  | 1   | 6   | 4    |
| CA4  | 0   | 0   | 0   | 1   | 1   | 1   | 1   | 1   | 5   | 3    |
| CA5  | 0   | 0   | 0   | 1   | 1   | 1   | 1   | 1   | 5   | 3    |
| CA6  | 0   | 0   | 0   | 1   | 1*  | 1   | 1*  | 1*  | 5   | 3    |
| CA7  | 0   | 0   | 0   | 0   | 0   | 0   | 1   | 1   | 2   | 1    |
| CA8  | 0   | 0   | 0   | 0   | 0   | 0   | 1   | 1   | 2   | 1    |
| Dep  | 2   | 1   | 2   | 5   | 5   | 4   | 8   | 8   | 35  |      |
| Rank | 2   | 1   | 2   | 4   | 4   | 3   | 5   | 5   |     |      |

### 3) Partitioning the reachability matrix

Warfield [69] defined the reachability set as an element itself that may help other elements to achieve certain objective. In contrast, the antecedent set includes the element and other elements that may help achieving it. The intersection of these sets is then derived from all the elements. The elements with the same reachability and intersection sets occupy the top level in the ISM hierarchy. The top-level element in the hierarchy does not help achieving any other element above its level. Once the top-level element is identified, it is separated from the other elements, and the same process is repeated to find out the elements at the next level. This process is continued until the level of each element is determined. It helps in building the digraph and the ISM model. Table IX presents the eight criteria we identified (key areas of AI ethics motivators), together with their reachability set, antecedent set, intersection set, and levels.

Table IX: Leveling of final reachability matrix

| | Reachability Set | Antecedent Set | Intersection Set | Level |
|---|---|---|---|---|
| Level Partitions | | | | |
| Iteration One | | | | |
| CA1 | 1,3,4,5,6,7,8 | 1,3 | 1,3 | |
| CA2 | 2,7,8 | 2 | 2 | |
| CA3 | 1,3,4,5,7,8 | 1,3 | 1,3 | |
| CA4 | 4,5,6,7,8 | 1,3,4,5,6 | 4,5,6 | |
| CA5 | 4,5,6,7,8 | 1,3,4,5,6 | 4,5,6 | |
| CA6 | 4,5,6,7,8 | 1,4,5,6 | 4,5,6 | |
| CA7 | 7,8 | 1,2,3,4,5,6,7,8 | 7,8 | level 1 |
| CA8 | 7,8 | 1,2,3,4,5,6,7,8 | 7,8 | level 1 |
| Iteration Two | | | | |
| CA1 | 1,3,4,5,6 | 1,3 | 1,3 | |
| CA2 | 2 | 2 | 2 | level 2 |
| CA3 | 1,3,4,5 | 1,3 | 1,3 | |
| CA4 | 4,5,6 | 1,3,4,5,6 | 4,5,6 | level 3 |
| CA5 | 4,5,6 | 1,3,4,5,6 | 4,5,6 | level 3 |
| CA6 | 4,5,6 | 1,4,5,6 | 4,5,6 | level 3 |
| Iteration Three | | | | |
| CA1 | 1,3 | 1,3 | 1,3 | level 4 |

*4) Interpretation of ISM model*

The final ISM model was developed using the reachability matrix results, which presented the inter-relationships of the criteria as arrows pointing from one criterion to another. To ensure data accuracy, transitivity analysis [70] was performed on the digraph once converted to the ISM model (see Fig 2). This analysis showed that CA1 (Human resource) and CA3 (Coordination) areas ranked at the top (level 4) as the AI ethics motivators. This indicates that CA1 and CA3 are independent areas of the AI ethics motivators, but there is an interrelationship between them. This suggests that both areas must support each other while developing AI-based systems and considering ethics. Based on the results presented in Fig 2, it was found that CA4 (Project administration), CA5 (Standards), and CA6 (Technology factor) are dependent only on the key-areas of Level 4, i.e., CA1 (Human resource) and CA3 (Coordination). At the same time, the rest of the key categories of AI ethics are dependent on them. This indicates that to implement AI ethics of Level-2, CA2 (Knowledge integration) and Level 1 Stakeholders (CA7), and Strategy and matrices (CA8), it is necessary to implement the motivators of key categories lying on level 3 and level 4. To implement the motivators of key categories of level 3, it is required to implement the motivator of only level-4, i.e., CA1 (Human resource) and CA3 (Coordination). In summary, CA1 (Human resource) and CA3 (Coordination) are fully independent categories of AI ethics motivators, whereas Stakeholders (CA7) and Strategy and matrices (CA8) are fully dependent on all other AI ethics categories.

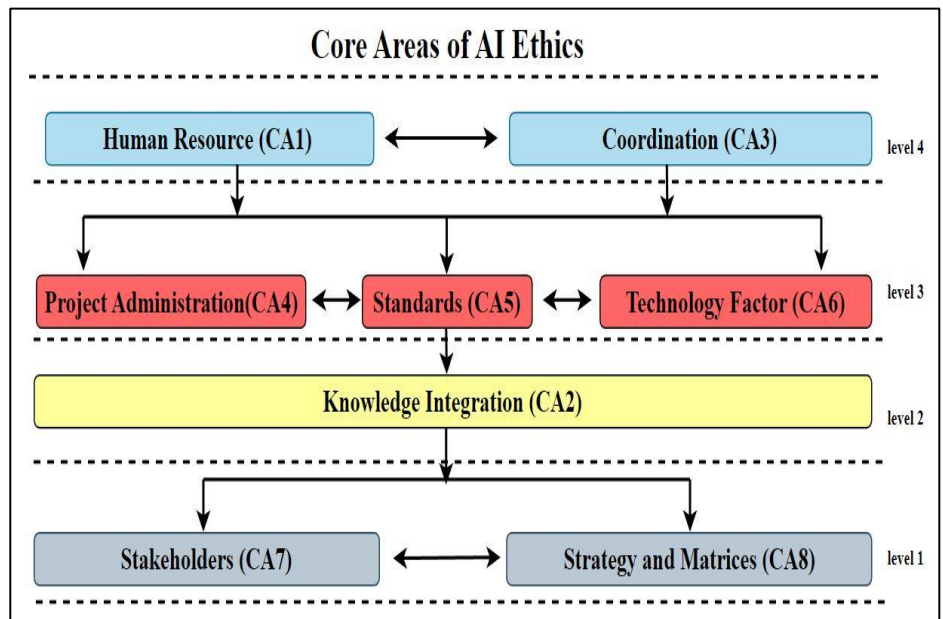

Fig 2: Leveling of AI ethics motivators core categories

### 5) MICMAC analysis

MICMAC is an acronym for Cross-Impact Matrix Multiplication Applied to Classification (in French: Matrice d'Impacts Croisés Multiplication Appliqueeaun Classement). It is useful for examining the key categories that drive a complex system. According to Attri et al. (72), MICMAC "is an analysis that examines categories' driving power and dependence power." Enablers are classified into four clusters based on their driving and dependence power. Hence, MICMAC is a useful tool for examining categories' driving power and dependence power, as well as classifying enablers into four distinct clusters listed below. Understanding these clusters can help identify key enablers and their impact on the system.

a) **Autonomous Cluster:** This cluster includes categories with weak driving and dependence power, often remaining disconnected from the system due to weak linkages. As a result, these categories have a minimal impact on the overall system.

b) **Linkage Cluster:** Categories within this cluster exhibit both strong driving and dependence power, significantly influencing other enablers due to their strong interconnections.

c) **Dependent Cluster:** Categories in this cluster possess strong dependence power but weak driving power, making them reliant on other factors within the system.

d) **Independent Cluster:** This cluster consists of categories with strong driving power but weak dependence power. These enablers, often referred to as "key motivators," play a crucial role in influencing the system.

### 6) Development of conical matrix

The key objective of conical matrix development was to perform the MICMAC analysis. The conical matrix presented in Table X, was developed considering the data from Tables VIII and IX. Firstly, all the criteria were ordered concerning their level number (Table IX). Secondly, the value of each criterion was considered, based on Table VIII. For example, the value of CA2 across rows and columns of the transitivity matrix (Table VIII) shows the '0' relationship between all categories except itself. Similarly, the value of C7 shows a '1' relationship with C1, C2, C3, C4, C5, C6, C7, and C8. Thus, considering the same procedure, our conical matrix is formulated in Table X.

Table X: Conical matrix after clustering enablers

|  | CA7 | CA8 | CA2 | CA4 | CA5 | CA6 | CA3 | CA1 |
|---|---|---|---|---|---|---|---|---|
| CA7 | 1 | 1 | 0 | 0 | 0 | 0 | 0 | 0 |
| CA8 | 1 | 1 | 0 | 0 | 0 | 0 | 0 | 0 |
| CA2 | 1 | 1 | 1 | 0 | 0 | 0 | 0 | 0 |
| CA4 | 1 | 1 | 0 | 1 | 1 | 1 | 0 | 0 |
| CA5 | 1 | 1 | 0 | 1 | 1 | 1 | 0 | 0 |
| CA6 | 1 | 1 | 0 | 1 | 1 | 1 | 0 | 0 |
| CA3 | 1 | 1 | 0 | 1 | 1 | 0 | 1 | 1 |
| CA1 | 1 | 1 | 0 | 1 | 1 | 1 | 1 | 1 |

Then, we used the classification approach of Kannan et al. [53] to classify the results of the MICMAC analysis, which are presented in Figure 3. The MICMAC analysis classified all the AI ethics categories into four clusters. Figure 3 shows the classification of the AI ethics core categories into these four clusters. The first cluster includes the autonomous categories of AI ethics motivators. The second cluster includes the dependent categories of AI ethics motivators, whereas the third and fourth clusters include the independent and linkage relationships of the categories of AI ethics motivators. Based on the results presented in Figure 3, the CA2 (Knowledge integration) category is part of the autonomous cluster. This indicates that CA2 is disconnected from the system due to weak links with other areas of AI ethics. The results show that Project administration (CA4) and Standards (CA5) belong to the linkage factors cluster, indicating that these categories have strong driving and dependence power, and affect other enablers due to strong linkages.

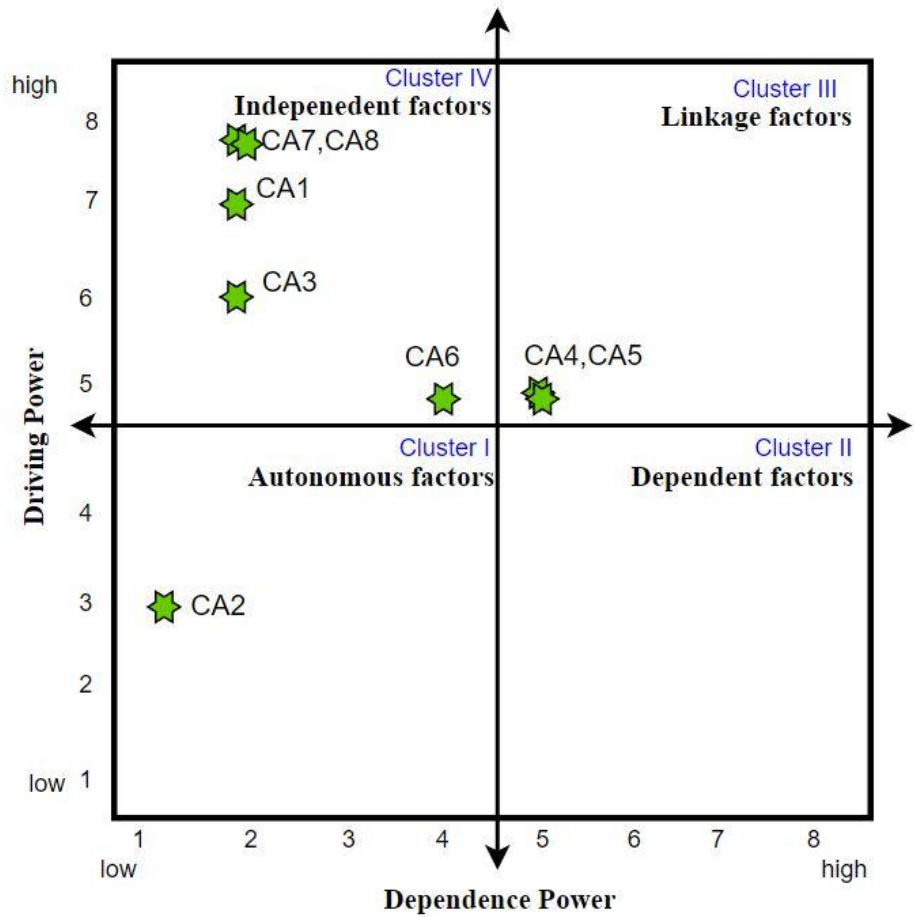

Fig 3: Graphical view of MICMAC analysis

Furthermore, we observed that CA1 (Human resource), CA3 (Coordination), CA7 (Stakeholders), and CA8 (Strategy and matrices) belong to the independent cluster. This suggests that CA1, CA3, CA7, and CA8 have weak dependence power but strong driving power. Interestingly, no category belongs to the fully dependent cluster. Finally, the results presented in Figure 3 provide a useful classification of the AI ethics categories into four clusters, which can help identify each category's driving and dependence power, as well as its impact on the system.

## C. Application of Fuzzy TOPSIS

Whereas our ISM analysis identified key categories of AI ethics motivators, it did not cover uncertainties or vagueness in considering these motivators. To address these issues, we applied the Fuzzy TOPSIS approach, which ranks challenging factors and assists practitioners in prioritizing the most effective motivators for considering AI ethics in information system development. The Fuzzy TOPSIS approach has already been used in other engineering domains to address multi-criteria decision-making problems [59, 71-73]. For this purpose, we requested ten experts who participated in our ISM analysis (Appendix C) to rank each motivator based on their experience and understanding, by using a questionnaire developed for this purpose (see Appendix C). Each participant was allowed to consult other colleagues whereas ranking the complex factors to ensure a representative response from each experts. We followed the below steps to calculate the Fuzzy TOPSIS results:

*1) Steps 1 & 2:* We used the Fuzzy Triangular Scale to obtain insights from experts regarding the effectiveness of AI ethics motivators, which gave linguistic values (Table IV) [53-55].

*2) Step 3:* We computed the combined decision matrix using (1) in Section III-C. We studied a total of 20 motivators related to the eight core categories of AI ethics. Table XI shows the combined decision matrix, which represents the collective opinion of all experts involved in decision-making.

### Table XI: Combined decision matrix

| | Human resource | | | Knowledge integration | | | Coordination | | | Project administration | | | Standards | | | Technology factor | | | Stakeholders | | | Strategy matrices | | |
| --- | --- | --- | --- | --- | --- | --- | --- | --- | --- | --- | --- | --- | --- | --- | --- | --- | --- | --- | --- | --- | --- | --- | --- | --- |
| WEIGHTS | 1 | 1,2 | 2 | 1,2 | 2,5 | 3,5 | 0,5 | 1 | 2 | 0,5 | 1,4 | 2,5 | 0,5 | 1 | 1,5 | 0,5 | 1,4 | 2,5 | 0,5 | 1,2 | 2 | 0,5 | 1,3 | 3 |
| M1 | 0,5 | 2,5 | 3,5 | 1,5 | 2,4 | 3,5 | 1,5 | 2,4 | 3,5 | 1 | 1,2 | 2,5 | 0,5 | 1,3 | 1,5 | 1,5 | 2,5 | 3,5 | 0,5 | 1,3 | 2,5 | 0,5 | 1,7 | 3 |
| M2 | 0,5 | 1,6 | 3 | 0,5 | 2,1 | 3 | 1,5 | 2 | 3 | 0,5 | 1,3 | 3 | 0,5 | 1,4 | 3 | 1,5 | 2,2 | 3,5 | 1,5 | 2,4 | 3,5 | 0,5 | 1,4 | 3,5 |
| M3 | 0,5 | 1,4 | 3 | 1 | 1,2 | 2,5 | 0,5 | 1,7 | 3 | 1,5 | 1,2 | 3 | 0,5 | 1,2 | 2,5 | 1,5 | 2,4 | 3,5 | 0,5 | 1 | 1,5 | 0,5 | 1 | 1,5 |
| M4 | 1,5 | 2,5 | 3 | 0,5 | 2,1 | 3 | 1,5 | 2,3 | 3,5 | 0,5 | 1,2 | 2,5 | 0,5 | 1,2 | 2,5 | 0,5 | 2,1 | 3 | 0,5 | 1,7 | 3 | 0,5 | 1 | 1,5 |
| M5 | 0,5 | 2 | 3 | 0,5 | 2,1 | 3 | 0,5 | 1,7 | 3 | 0,5 | 1,2 | 2,5 | 0,5 | 1 | 1,5 | 0,5 | 2,1 | 3 | 0,5 | 1,7 | 3 | 0,5 | 1,4 | 3,5 |
| M6 | 0,5 | 1,4 | 3 | 0,5 | 2,1 | 3 | 1,5 | 2,3 | 3,5 | 1,5 | 2,3 | 3 | 0,5 | 1 | 1,5 | 0,5 | 2,1 | 3 | 1 | 1,3 | 2 | 0,5 | 1 | 1,5 |
| M7 | 0,5 | 1,4 | 3 | 1,5 | 2,3 | 3 | 0,5 | 2,1 | 3,5 | 0,5 | 1,2 | 2,5 | 0,5 | 1 | 1,5 | 1,5 | 2,4 | 3,5 | 1 | 1,3 | 2 | 0,5 | 1 | 1,5 |
| M8 | 0,5 | 1,4 | 3 | 1 | 2 | 3 | 0,5 | 2,1 | 3,5 | 0,5 | 1,2 | 2,5 | 1,3 | 3 | 3,5 | 0,5 | 1,7 | 3 | 1 | 1,3 | 2 | 0,5 | 1 | 1,5 |
| M9 | 1,5 | 2,1 | 3 | 1 | 1,4 | 2 | 0,5 | 2,1 | 3,5 | 0,1 | 1,8 | 3 | 0,5 | 2 | 3 | 1,5 | 2,4 | 3,5 | 0,5 | 1 | 1,5 | 0,5 | 2 | 3 |
| M10 | 1,5 | 2,1 | 3 | 0,5 | 1,4 | 2,5 | 1,5 | 2,4 | 3,5 | 1 | 1,4 | 2 | 0,5 | 1 | 1,5 | 0,5 | 2,1 | 3 | 0,5 | 1 | 2 | 0,5 | 1 | 1,5 |
| M11 | 1,5 | 2,1 | 3 | 1 | 1,4 | 2 | 1,5 | 2,4 | 3,5 | 0,1 | 1,8 | 3 | 0,5 | 1,3 | 2 | 0,5 | 2,1 | 3 | 1 | 1,3 | 2 | 0,5 | 1,2 | 2,5 |
| M12 | 0,5 | 1 | 1,5 | 1 | 1,4 | 2 | 1,5 | 2,3 | 3,5 | 1 | 1,4 | 2 | 0,5 | 1,3 | 2 | 1,5 | 2,3 | 3 | 1,5 | 1 | 1,5 | 0,5 | 2 | 3 |
| M13 | 0,5 | 1 | 1,5 | 0,5 | 1,4 | 2,5 | 0,5 | 2,1 | 3,5 | 0,1 | 1,8 | 3 | 0,5 | 1,3 | 2 | 0,5 | 1 | 1,5 | 0,5 | 1 | 2 | 0,5 | 1 | 1,5 |
| M14 | 1,5 | 2,5 | 3 | 1 | 1,4 | 2 | 0,5 | 2,1 | 3,5 | 0,1 | 1,8 | 2 | 0,5 | 1 | 1,5 | 0,5 | 1,7 | 3 | 1 | 1,3 | 2 | 0,5 | 1,2 | 2,5 |
| M15 | 0,5 | 2 | 3 | 0,5 | 2,1 | 3 | 1,5 | 2,4 | 3,5 | 0,1 | 1,8 | 3 | 1,3 | 3 | 3,5 | 1,5 | 2,4 | 3,5 | 0,5 | 1 | 1,5 | 0,5 | 1,4 | 3,5 |
| M16 | 0,5 | 1,4 | 3 | 1,5 | 2,3 | 3 | 1,5 | 2,4 | 3,5 | 1 | 1,4 | 2 | 0,5 | 2 | 3 | 0,5 | 2,1 | 3 | 0,5 | 1,7 | 3 | 0,5 | 1 | 1,5 |
| M17 | 0,5 | 1,4 | 3 | 1 | 2 | 3 | 0,5 | 1,7 | 3 | 0,1 | 1,8 | 3 | 0,5 | 1 | 1,5 | 0,5 | 1,7 | 3 | 0,5 | 1,7 | 3 | 0,5 | 1,4 | 3,5 |
| M18 | 1,5 | 2,1 | 3 | 1 | 1,4 | 2 | 1,5 | 2,3 | 3,5 | 1 | 1,4 | 2 | 1,3 | 3 | 3,5 | 1,5 | 2,4 | 3,5 | 1 | 1,3 | 2 | 0,5 | 1 | 1,5 |
| M19 | 0,5 | 1 | 1,5 | 0,5 | 1,4 | 2,5 | 0,5 | 1,7 | 3 | 1,5 | 2,3 | 3 | 0,5 | 2 | 3 | 0,5 | 1,7 | 3 | 0,5 | 1 | 1,5 | 0,5 | 2 | 3 |
| M20 | 0,5 | 1 | 1,5 | 0,1 | 1,8 | 2 | 1,5 | 2,3 | 3,5 | 1,5 | 2,3 | 3 | 0,5 | 1 | 1,5 | 1,5 | 2,4 | 3,5 | 0,5 | 1 | 1,5 | 0,5 | 2 | 3 |

*3) Step 4:* At this stage, we calculated the normalized decision matrix using (2) and (3) in section III-C. The normalization process considers both cost and benefit criteria, which is a systematic approach to evaluate the strengths and weaknesses of alternatives when determining the best approach for achieving benefits, whilst performing a specific task [74]. To normalize the decision matrix, we utilized 'Human Resources' as a cost criterion, which included motivators such as 'Building diverse teams' and 'Assigning a leader responsible and accountable for ethical AI'. The results of the normalization process are presented in Table XII."

Table XII: Normalized decision matrix

| WEIGHTS | Human Resources | | | Knowledge Integration | | | Coordination | | | Project Administration | | | Standards | | | Technology Factor | | | Stakeholders | | | Strategy Metrics | | |
|---|---|---|---|---|---|---|---|---|---|---|---|---|---|---|---|---|---|---|---|---|---|---|---|---|
| | 1 | 1,2 | 2 | 1,2 | 2,5 | 3,5 | 0,5 | 0,5 | 0,5 | 1 | 1,4 | 2,5 | 0,5 | 1 | 1,5 | 0,5 | 1,4 | 2,5 | 0,5 | 1,2 | 2 | 0,5 | 1,3 | 3 |
| M1 | 0,143 | 0,200 | 1,000 | 0,429 | 0,686 | 1,000 | 0,429 | 0,686 | 1,000 | 0,333 | 0,400 | 0,833 | 0,143 | 0,371 | 0,429 | 0,429 | 0,714 | 1,000 | 0,143 | 0,371 | 0,714 | 0,143 | 0,486 | 0,857 |
| M2 | 0,167 | 0,313 | 1,000 | 0,143 | 0,600 | 0,857 | 0,429 | 0,571 | 0,857 | 0,167 | 0,433 | 1,000 | 0,143 | 0,400 | 0,857 | 0,429 | 0,629 | 1,000 | 0,429 | 0,686 | 1,000 | 0,143 | 0,400 | 1,000 |
| M3 | 0,167 | 0,357 | 1,000 | 0,286 | 0,343 | 0,714 | 0,143 | 0,486 | 0,857 | 0,500 | 0,400 | 1,000 | 0,143 | 0,343 | 0,714 | 0,429 | 0,686 | 1,000 | 0,143 | 0,286 | 0,429 | 0,143 | 0,286 | 0,429 |
| M4 | 0,167 | 0,200 | 0,333 | 0,143 | 0,600 | 0,857 | 0,429 | 0,657 | 1,000 | 0,167 | 0,400 | 0,833 | 0,143 | 0,343 | 0,714 | 0,143 | 0,600 | 0,857 | 0,143 | 0,486 | 0,857 | 0,143 | 0,286 | 0,429 |
| M5 | 0,167 | 0,250 | 1,000 | 0,143 | 0,600 | 0,857 | 0,143 | 0,486 | 0,857 | 0,167 | 0,400 | 0,833 | 0,143 | 0,286 | 0,429 | 0,143 | 0,600 | 0,857 | 0,143 | 0,486 | 0,857 | 0,143 | 0,400 | 1,000 |
| M6 | 0,167 | 0,357 | 1,000 | 0,143 | 0,600 | 0,857 | 0,429 | 0,657 | 1,000 | 0,500 | 0,767 | 1,000 | 0,143 | 0,286 | 0,429 | 0,143 | 0,600 | 0,857 | 0,286 | 0,371 | 0,571 | 0,143 | 0,286 | 0,429 |
| M7 | 0,167 | 0,357 | 1,000 | 0,429 | 0,657 | 0,857 | 0,143 | 0,600 | 1,000 | 0,167 | 0,400 | 0,833 | 0,143 | 0,286 | 0,429 | 0,429 | 0,686 | 1,000 | 0,286 | 0,371 | 0,571 | 0,143 | 0,286 | 0,429 |
| M8 | 0,167 | 0,357 | 1,000 | 0,286 | 0,571 | 0,857 | 0,143 | 0,600 | 1,000 | 0,167 | 0,400 | 0,833 | 0,371 | 0,857 | 1,000 | 0,429 | 0,686 | 0,857 | 0,286 | 0,371 | 0,571 | 0,143 | 0,286 | 0,429 |
| M9 | 0,167 | 0,238 | 0,333 | 0,286 | 0,400 | 0,571 | 0,143 | 0,600 | 1,000 | 0,033 | 0,600 | 1,000 | 0,143 | 0,571 | 0,857 | 0,429 | 0,686 | 1,000 | 0,143 | 0,286 | 0,429 | 0,143 | 0,571 | 0,857 |
| M10 | 0,167 | 0,238 | 0,333 | 0,143 | 0,400 | 0,714 | 0,429 | 0,686 | 1,000 | 0,333 | 0,467 | 0,667 | 0,143 | 0,286 | 0,429 | 0,143 | 0,600 | 0,857 | 0,143 | 0,486 | 0,857 | 0,143 | 0,286 | 0,429 |
| M11 | 0,167 | 0,238 | 0,333 | 0,286 | 0,400 | 0,571 | 0,429 | 0,686 | 1,000 | 0,033 | 0,600 | 1,000 | 0,143 | 0,371 | 0,571 | 0,143 | 0,600 | 0,857 | 0,286 | 0,371 | 0,571 | 0,143 | 0,343 | 0,714 |
| M12 | 0,333 | 0,500 | 1,000 | 0,286 | 0,400 | 0,571 | 0,429 | 0,657 | 1,000 | 0,333 | 0,467 | 0,667 | 0,143 | 0,400 | 0,857 | 0,371 | 0,571 | 0,429 | 0,657 | 0,143 | 0,486 | 0,857 | 0,143 | 0,286 |
| M13 | 0,333 | 0,500 | 1,000 | 0,143 | 0,400 | 0,714 | 0,143 | 0,600 | 1,000 | 0,033 | 0,600 | 1,000 | 0,143 | 0,371 | 0,429 | 0,143 | 0,286 | 0,429 | 0,143 | 0,286 | 0,429 | 0,143 | 0,286 | 0,429 |
| M14 | 0,167 | 0,200 | 0,333 | 0,286 | 0,400 | 0,714 | 0,143 | 0,600 | 1,000 | 0,600 | 0,667 | 1,000 | 0,143 | 0,286 | 0,429 | 0,143 | 0,486 | 0,857 | 0,286 | 0,371 | 0,571 | 0,143 | 0,343 | 0,714 |
| M15 | 0,167 | 0,250 | 1,000 | 0,143 | 0,600 | 0,857 | 0,429 | 0,686 | 1,000 | 0,143 | 0,600 | 0,833 | 0,143 | 0,371 | 0,857 | 0,429 | 0,686 | 1,000 | 0,143 | 0,286 | 0,429 | 0,143 | 0,400 | 1,000 |
| M16 | 0,167 | 0,357 | 1,000 | 0,429 | 0,657 | 0,857 | 0,429 | 0,686 | 1,000 | 0,167 | 0,653 | 0,667 | 0,143 | 0,571 | 0,857 | 0,143 | 0,486 | 0,857 | 0,143 | 0,286 | 0,429 | 0,143 | 0,286 | 0,429 |
| M17 | 0,167 | 0,357 | 1,000 | 0,286 | 0,571 | 0,857 | 0,143 | 0,486 | 0,857 | 0,033 | 0,600 | 1,000 | 0,143 | 0,286 | 0,429 | 0,429 | 0,686 | 1,000 | 0,143 | 0,486 | 0,857 | 0,143 | 0,400 | 1,000 |
| M18 | 0,167 | 0,238 | 0,333 | 0,286 | 0,400 | 0,571 | 0,429 | 0,657 | 1,000 | 0,167 | 0,653 | 0,667 | 0,371 | 0,857 | 1,000 | 0,429 | 0,686 | 1,000 | 0,286 | 0,371 | 0,571 | 0,143 | 0,286 | 0,429 |
| M19 | 0,333 | 0,500 | 1,000 | 0,143 | 0,400 | 0,714 | 0,143 | 0,486 | 0,857 | 0,500 | 0,767 | 1,000 | 0,143 | 0,571 | 0,857 | 0,143 | 0,486 | 0,857 | 0,143 | 0,486 | 0,857 | 0,143 | 0,571 | 0,857 |
| M20 | 0,333 | 0,500 | 1,000 | 0,029 | 0,514 | 0,571 | 0,429 | 0,657 | 1,000 | 0,500 | 0,767 | 1,000 | 0,143 | 0,286 | 0,429 | 0,429 | 0,686 | 1,000 | 0,143 | 0,286 | 0,429 | 0,143 | 0,571 | 0,857 |

*4) Step 5:* We calculated the weighted decision matrix in this stage by multiplying the weights assigned by the group of experts to each criterion with their corresponding alternative (i.e., motivators) using equation 4. The resulting values are presented in Table XIII.

Table XIII: Weighted normalized decision matrix

| Weights | Human resources | | | Knowledge integration | | | Coordination | | | Project administration | | | Standards | | | Technology factor | | | Stakeholders | | | Strategy metrics | | |
|---|---|---|---|---|---|---|---|---|---|---|---|---|---|---|---|---|---|---|---|---|---|---|---|---|
| | 1 | 1,2 | 2 | 1,2 | 2,5 | 3,5 | 0,5 | 0,5 | 0,5 | 1 | 1,4 | 2,5 | 0,5 | 1 | 1,5 | 0,5 | 1,4 | 2,5 | 0,5 | 1,2 | 2 | 0,5 | 1,3 | 3 |
| M1 | 0,143 | 0,240 | 2,000 | 0,514 | 1,714 | 3,500 | 0,214 | 0,686 | 2,000 | 0,167 | 0,560 | 2,083 | 0,071 | 0,371 | 0,643 | 0,214 | 1,000 | 2,500 | 0,071 | 0,446 | 1,429 | 0,071 | 0,631 | 2,571 |
| M2 | 0,167 | 0,375 | 2,000 | 0,171 | 1,500 | 3,000 | 0,214 | 0,571 | 1,714 | 0,083 | 0,607 | 2,500 | 0,071 | 0,400 | 1,286 | 0,214 | 0,880 | 2,500 | 0,214 | 0,823 | 2,000 | 0,071 | 0,520 | 3,000 |
| M3 | 0,167 | 0,429 | 2,000 | 0,343 | 0,857 | 2,500 | 0,071 | 0,486 | 1,714 | 0,250 | 0,560 | 2,500 | 0,071 | 0,343 | 1,071 | 0,214 | 0,960 | 2,500 | 0,071 | 0,343 | 0,857 | 0,071 | 0,371 | 1,286 |
| M4 | 0,167 | 0,240 | 0,667 | 0,171 | 1,500 | 3,000 | 0,214 | 0,657 | 2,000 | 0,083 | 0,560 | 2,083 | 0,071 | 0,343 | 0,714 | 0,071 | 0,840 | 2,143 | 0,071 | 0,583 | 1,714 | 0,071 | 0,371 | 1,286 |
| M5 | 0,167 | 0,300 | 2,000 | 0,171 | 1,500 | 3,000 | 0,071 | 0,486 | 1,714 | 0,083 | 0,560 | 2,083 | 0,071 | 0,286 | 0,643 | 0,071 | 0,840 | 2,143 | 0,071 | 0,583 | 1,714 | 0,071 | 0,520 | 3,000 |
| M6 | 0,167 | 0,429 | 2,000 | 0,171 | 1,500 | 3,000 | 0,214 | 0,657 | 2,000 | 0,250 | 1,073 | 2,500 | 0,071 | 0,286 | 0,643 | 0,071 | 0,840 | 2,143 | 0,143 | 0,446 | 1,143 | 0,071 | 0,371 | 1,286 |
| M7 | 0,167 | 0,429 | 2,000 | 0,514 | 1,643 | 3,000 | 0,071 | 0,600 | 2,000 | 0,083 | 0,560 | 2,083 | 0,071 | 0,286 | 0,643 | 0,214 | 0,960 | 2,500 | 0,143 | 0,446 | 1,143 | 0,071 | 0,371 | 1,286 |
| M8 | 0,167 | 0,429 | 2,000 | 0,343 | 1,429 | 3,000 | 0,071 | 0,600 | 2,000 | 0,083 | 0,560 | 2,083 | 0,186 | 0,857 | 1,500 | 0,071 | 0,680 | 2,143 | 0,143 | 0,446 | 1,143 | 0,071 | 0,371 | 1,286 |
| M9 | 0,167 | 0,286 | 0,667 | 0,343 | 1,000 | 2,000 | 0,071 | 0,600 | 2,000 | 0,017 | 0,840 | 2,500 | 0,071 | 0,571 | 1,286 | 0,214 | 0,960 | 2,500 | 0,071 | 0,343 | 0,857 | 0,071 | 0,743 | 2,571 |
| M10 | 0,167 | 0,286 | 0,667 | 0,171 | 1,000 | 2,500 | 0,214 | 0,686 | 2,000 | 0,167 | 0,653 | 1,667 | 0,071 | 0,286 | 0,643 | 0,071 | 0,840 | 2,143 | 0,071 | 0,583 | 1,714 | 0,071 | 0,371 | 1,286 |
| M11 | 0,167 | 0,286 | 0,667 | 0,343 | 1,000 | 2,000 | 0,214 | 0,686 | 2,000 | 0,017 | 0,840 | 2,500 | 0,071 | 0,371 | 0,857 | 0,071 | 0,840 | 2,143 | 0,143 | 0,446 | 1,143 | 0,071 | 0,446 | 2,143 |
| M12 | 0,333 | 0,600 | 2,000 | 0,343 | 1,000 | 2,000 | 0,214 | 0,657 | 2,000 | 0,167 | 0,653 | 1,667 | 0,071 | 0,371 | 0,857 | 0,214 | 0,920 | 2,143 | 0,071 | 1,071 | 0,071 | 0,343 | 0,743 | 2,571 |
| M13 | 0,333 | 0,600 | 2,000 | 0,171 | 1,000 | 2,500 | 0,071 | 0,600 | 2,000 | 0,017 | 0,840 | 2,500 | 0,071 | 0,371 | 0,857 | 0,071 | 0,857 | 2,143 | 0,143 | 0,583 | 2,000 | 0,071 | 0,520 | 3,000 |
| M14 | 0,167 | 0,240 | 0,667 | 0,343 | 1,000 | 2,000 | 0,071 | 0,600 | 2,000 | 0,017 | 0,840 | 2,500 | 0,071 | 0,286 | 0,643 | 0,071 | 0,680 | 2,143 | 0,143 | 0,446 | 2,143 | 0,071 | 0,371 | 1,286 |
| M15 | 0,167 | 0,300 | 2,000 | 0,171 | 1,500 | 3,000 | 0,071 | 0,486 | 1,714 | 0,083 | 0,560 | 2,083 | 0,071 | 0,286 | 0,643 | 0,071 | 0,840 | 2,143 | 0,143 | 0,583 | 0,857 | 0,071 | 0,520 | 3,000 |
| M16 | 0,167 | 0,429 | 2,000 | 0,514 | 1,643 | 3,000 | 0,071 | 0,600 | 2,000 | 0,167 | 0,653 | 1,667 | 0,071 | 0,571 | 1,286 | 0,071 | 0,680 | 2,143 | 0,143 | 0,583 | 1,714 | 0,071 | 0,371 | 1,286 |
| M17 | 0,167 | 0,429 | 2,000 | 0,343 | 1,429 | 3,000 | 0,071 | 0,486 | 1,714 | 0,083 | 0,560 | 2,500 | 0,071 | 0,286 | 0,643 | 0,214 | 0,960 | 2,500 | 0,071 | 0,583 | 1,714 | 0,071 | 0,520 | 3,000 |
| M18 | 0,167 | 0,286 | 0,667 | 0,343 | 1,000 | 2,000 | 0,214 | 0,657 | 2,000 | 0,167 | 0,653 | 1,667 | 0,186 | 0,857 | 1,500 | 0,214 | 0,960 | 2,500 | 0,143 | 0,446 | 1,143 | 0,071 | 0,371 | 1,286 |
| M19 | 0,333 | 0,600 | 2,000 | 0,171 | 1,000 | 2,500 | 0,071 | 0,486 | 1,714 | 0,250 | 1,073 | 2,500 | 0,071 | 0,571 | 1,286 | 0,071 | 0,680 | 2,143 | 0,143 | 0,583 | 0,857 | 0,071 | 0,743 | 2,571 |
| M20 | 0,333 | 0,500 | 1,000 | 0,034 | 1,286 | 2,000 | 0,214 | 0,657 | 2,000 | 0,250 | 1,073 | 2,500 | 0,071 | 0,286 | 0,643 | 0,214 | 0,960 | 2,500 | 0,071 | 0,343 | 0,857 | 0,071 | 0,743 | 2,571 |
| A+ | 0,333 | 0,600 | 2,000 | 0,514 | 1,714 | 3,500 | 0,214 | 0,686 | 2,000 | 0,250 | 1,073 | 2,500 | 0,186 | 0,857 | 1,500 | 0,214 | 1,000 | 2,500 | 0,214 | 0,823 | 2,000 | 0,071 | 0,743 | 3,000 |
| A- | 0,143 | 0,240 | 0,667 | 0,034 | 0,857 | 2,000 | 0,071 | 0,486 | 1,714 | 0,017 | 0,560 | 1,667 | 0,071 | 0,286 | 0,643 | 0,071 | 0,400 | 1,071 | 0,071 | 0,343 | 0,857 | 0,071 | 0,371 | 1,286 |

*5) Step 6:* We calculated the fuzzy ideal solutions (positive and negative) by executing equations 5 and 6 at this stage. The resulting fuzzy positive and negative solutions can be found in Tables XIV and XV.

*6) Step 7:* We then determined distance values (di* and di-) for each ai ethics motivator for the criteria using equations 7 and 8. The results are presented in tables xiv and xv. The core categories for all AI ethics motivators are CA1 to CA8.

Table XIV: Fuzzy positive ideal solution

| | CA1 | CA2 | CA3 | CA4 | CA5 | CA6 | CA7 | CA8 | DI* |
|---|---|---|---|---|---|---|---|---|---|
| M1 | 0,235 | 0,000 | 0,000 | 0,385 | 0,573 | 0,000 | 0,404 | 0,256 | 1,852 |
| M2 | 0,162 | 0,371 | 0,178 | 0,286 | 0,299 | 0,069 | 0,000 | 0,129 | 1,493 |
| M3 | 0,138 | 0,767 | 0,218 | 0,296 | 0,392 | 0,023 | 0,720 | 1,013 | 3,567 |
| M4 | 0,803 | 0,371 | 0,016 | 0,394 | 0,392 | 0,241 | 0,231 | 1,013 | 3,461 |
| M5 | 0,198 | 0,371 | 0,218 | 0,394 | 0,598 | 0,241 | 0,231 | 0,129 | 2,379 |
| M6 | 0,138 | 0,371 | 0,016 | 0,000 | 0,598 | 0,241 | 0,542 | 1,013 | 2,920 |
| M7 | 0,138 | 0,292 | 0,096 | 0,394 | 0,598 | 0,023 | 0,542 | 1,013 | 3,096 |
| M8 | 0,138 | 0,347 | 0,096 | 0,394 | 0,000 | 0,289 | 0,542 | 1,013 | 2,819 |
| M9 | 0,797 | 0,964 | 0,096 | 0,191 | 0,216 | 0,023 | 0,720 | 0,247 | 3,255 |
| M10 | 0,797 | 0,737 | 0,000 | 0,541 | 0,598 | 0,241 | 0,573 | 1,013 | 4,499 |
| M11 | 0,797 | 0,964 | 0,000 | 0,191 | 0,470 | 0,241 | 0,542 | 0,524 | 3,728 |
| M12 | 0,000 | 0,964 | 0,016 | 0,541 | 0,470 | 0,211 | 0,720 | 0,247 | 3,171 |
| M13 | 0,000 | 0,737 | 0,096 | 0,191 | 0,470 | 0,898 | 0,573 | 1,013 | 3,977 |
| M14 | 0,803 | 0,964 | 0,096 | 0,517 | 0,598 | 0,289 | 0,542 | 0,524 | 4,334 |
| M15 | 0,198 | 0,371 | 0,000 | 0,191 | 0,000 | 0,023 | 0,720 | 0,129 | 1,632 |
| M16 | 0,138 | 0,292 | 0,000 | 0,541 | 0,216 | 0,241 | 0,231 | 1,013 | 2,671 |

| | | | | | | | | |
|---|---|---|---|---|---|---|---|---|
| M17 | 0,138 | 0,347 | 0,218 | 0,191 | 0,598 | 0,289 | 0,231 | 0,129 | 2,140 |
| M18 | 0,797 | 0,964 | 0,016 | 0,541 | 0,000 | 0,023 | 0,542 | 1,013 | 3,896 |
| M19 | 0,000 | 0,737 | 0,218 | 0,000 | 0,216 | 0,289 | 0,720 | 0,247 | 2,427 |
| M20 | 0,000 | 0,942 | 0,016 | 0,000 | 0,598 | 0,023 | 0,720 | 0,247 | 2,548 |

Table XV: Fuzzy negative ideal solution

| | CA1 | CA2 | CA3 | CA4 | CA5 | CA6 | CA7 | CA8 | DI- |
|---|---|---|---|---|---|---|---|---|---|
| M1 | 0,770 | 1,035 | 0,218 | 0,256 | 0,049 | 0,898 | 0,335 | 0,757 | 1,035 |
| M2 | 0,774 | 0,691 | 0,096 | 0,483 | 0,377 | 0,874 | 0,720 | 0,993 | 0,993 |
| M3 | 0,778 | 0,339 | 0,000 | 0,500 | 0,250 | 0,890 | 0,000 | 0,000 | 0,890 |
| M4 | 0,014 | 0,691 | 0,209 | 0,244 | 0,250 | 0,669 | 0,514 | 0,000 | 0,691 |
| M5 | 0,771 | 0,691 | 0,000 | 0,244 | 0,000 | 0,669 | 0,514 | 0,993 | 0,993 |
| M6 | 0,778 | 0,691 | 0,209 | 0,581 | 0,000 | 0,669 | 0,180 | 0,000 | 0,778 |
| M7 | 0,778 | 0,785 | 0,178 | 0,244 | 0,000 | 0,890 | 0,180 | 0,000 | 0,890 |
| M8 | 0,778 | 0,688 | 0,178 | 0,244 | 0,598 | 0,639 | 0,180 | 0,000 | 0,778 |
| M9 | 0,030 | 0,196 | 0,178 | 0,508 | 0,406 | 0,890 | 0,000 | 0,773 | 0,890 |
| M10 | 0,030 | 0,310 | 0,218 | 0,102 | 0,000 | 0,669 | 0,165 | 0,000 | 0,669 |
| M11 | 0,030 | 0,196 | 0,218 | 0,508 | 0,133 | 0,669 | 0,180 | 0,497 | 0,669 |
| M12 | 0,805 | 0,196 | 0,209 | 0,102 | 0,133 | 0,693 | 0,000 | 0,773 | 0,805 |
| M13 | 0,805 | 0,310 | 0,178 | 0,508 | 0,133 | 0,000 | 0,165 | 0,000 | 0,805 |
| M14 | 0,014 | 0,196 | 0,178 | 0,162 | 0,000 | 0,639 | 0,180 | 0,497 | 0,639 |
| M15 | 0,771 | 0,691 | 0,218 | 0,508 | 0,598 | 0,890 | 0,000 | 0,993 | 0,993 |
| M16 | 0,778 | 0,785 | 0,218 | 0,102 | 0,406 | 0,669 | 0,514 | 0,000 | 0,785 |
| M17 | 0,778 | 0,688 | 0,000 | 0,508 | 0,000 | 0,639 | 0,514 | 0,993 | 0,993 |
| M18 | 0,030 | 0,196 | 0,209 | 0,102 | 0,598 | 0,890 | 0,180 | 0,000 | 0,890 |
| M19 | 0,805 | 0,310 | 0,000 | 0,581 | 0,406 | 0,639 | 0,000 | 0,773 | 0,805 |
| M20 | 0,805 | 0,247 | 0,209 | 0,581 | 0,000 | 0,890 | 0,000 | 0,773 | 0,890 |

*7) Step 8:* In Step 8, the closeness coefficient was computed using equation 9, and the results are presented in Table XVI.

Table XVI: Closeness coefficient values and Ranks

| Core Areas | Sr.# | Motivators | CCi | Local Ranks | Global Ranks |
|---|---|---|---|---|---|
| | M1 | Assign a leader responsible and accountable for ethical AI | 0.359 | 2 | 3 |
| Human Resource (CA1) | M2 | Build diverse teams | 0.399 | 1 | 1 |
| Knowledge | M3 | Education and awareness to foster an ethical mind-set | 0.200 | 2 | 14 |
| Integration (CA2) | M12 | Consider ethics by design | 0.202 | 1 | 13 |
| | M14 | Erosion of privacy | 0.129 | 2 | 20 |
| Coordination (CA3) | M19 | Two-way communication with users | 0.249 | 1 | 7 |
| | M5 | Surveillance practices for data gathering and privacy of court users | 0.295 | 1 | 5 |
| Project | M10 | Establish AI ethics scope according to business domain | 0.129 | 3 | 19 |
| Administration (CA4) | M7 | Regulatory compliance and standardization | 0.223 | 2 | 9 |
| | M6 | Requirements of trustworthy AI | 0.210 | 3 | 12 |
| | M9 | Develop ethics and privacy roadmap | 0.215 | 2 | 11 |
| Standards (CA5) | M17 | Controllable AI with clear accountability | 0.317 | 1 | 4 |
| | M8 | Algorithmic accountability | 0.216 | 1 | 10 |
| Technology | M13 | Automating ethics | 0.168 | 3 | 16 |
| Factor (CA6) | M18 | Optimize guidance and tools | 0.186 | 2 | 15 |
| | M4 | Ethics council development | 0.166 | 2 | 17 |
| Stakeholders (CA7) | M15 | Set up a governance body to implement measures of accountability | 0.378 | 1 | 2 |
| | M20 | AI collides with patent law | 0.259 | 1 | 6 |
| Strategy and | M16 | Screen the data used to train the AI system for bias | 0.227 | 2 | 8 |
| Metrics (CA8) | M11 | AI ethics risk assessment policies | 0.152 | 3 | 18 |

*8) Step 9:* The closeness coefficient values were used to determine the motivators' local and global ranks. A higher CCi value indicates a higher ranking for a motivator (see Table XVI for rankings). Local ranks indicate the priority order of a motivator within its core category. For example, within the human resource category (CA1), M1 (Assign a leader responsible and accountable for ethical AI, CCi 0.359) is ranked as the second, and M2 (Build diverse teams, CCi 0.399) is ranked as the first. Practitioners can use local ranks to prioritize motivators related to their working domains. Global ranks were also determined to prioritize motivators across all key categories. Such ranking can help project managers or C-level managers with reviewing and creating new strategies for ethics in AI-based software. Results in Table XVI show that M2 (Build diverse teams) is the highest priority motivator among the twenty AI ethics motivators considered. Building diverse teams has significant importance while considering ethics in AI-based software development. Lennart *et al.* [75] and Janice *et al.*[76] suggested that a diverse team with different backgrounds, perspectives, and experiences could bring a broader range of ideas, viewpoints, and solutions to ethical dilemmas in AI. By bringing together individuals with diverse skill sets and knowledge, the team can ensure that their software is inclusive and fair for everyone. Moreover, a diverse team is able to identify and address potential biases in AI systems, ensuring that the software does not discriminate anyone against any group. Additionally, having a diverse team can help to foster a culture of ethical awareness and accountability within the development process [77]. By promoting an inclusive environment, the team can ensure that ethical considerations are taken into account throughout the software development lifecycle [78]. Therefore, building diverse teams is crucial for developing ethical AI-based software that meets the needs and values of a diverse society. Aside that, M15 (Set up a governance body to implement accountability measures) is ranked as the 2nd highest important motivator for considering ethics in AI-based software development. For example, World Health Organization [79] indicated that setting up a governance body is crucial in ensuring that ethical standards are incorporated into AI-based software development. The governance body should be responsible for establishing guidelines and procedures that ensure ethical practices throughout the software development process. By implementing accountability measures, the governance body can hold developers and stakeholders responsible for meeting ethical standards [80]. This can involve developing codes of conduct and ethical standards that guide the development of AI systems. The governance body can also play a crucial role in ensuring that the AI-based software is transparent and explainable, at the same time adhering to ethical principles such as privacy, fairness, and non-discrimination [81]. Additionally, the governance body should oversee the auditing and testing of the software to ensure that it is free from biases and conforms to ethical standards [82]. Ultimately, establishing a governance body can help fostering a culture of ethical responsibility and accountability within the organization, ensuring that ethical considerations are integrated into every aspect of AI-based software development. What is more, we further noted that M1 (Assign a leader responsible and accountable for ethical AI), M17 (Controllable AI with clear accountability), and M5 (Surveillance practices for data gathering and privacy of court users) are ranked as 3rd, 4th, and 5th most important motivators for the consideration of ethics in AI-based systems.

### D. Holistic model of AI ethics motivators

The holistic model for AI ethics motivators was developed using core categories and their respective motivators. To create this model, we utilized the results of the ISM approach to level the core categories of AI ethics motivators. At the same time, the fuzzy TOPSIS method was used to prioritize the motivators based on their significance for considering ethical motivators during AI-based software development. The leveling of the core categories revealed the internal relationships (dependency and independency) among them. At the same time, the priority order showed the significance of each motivator in addressing specific core areas of AI ethics. Local ranking was used to prioritized motivators within their core category, whereas global ranking allowed to compare motivators across all categories. The results presented in Figure 4 indicate that the human resource and coordination categories (CA1 and CA3) are at level 4 and fully independent but have strong driving power. The motivators "Assign a leader responsible and accountable for ethical AI" (M1) and "Build diverse teams" (M2) are part of the human resource category and are ranked as the 3rd and 1st most significant motivators for ethics consideration in AI-based system development, respectively. On the other hand, the motivators "Erosion of privacy" (M14) and "Two-way communication with users" (M19) are part of the coordination category (CA3), and they are ranked as the 20th and 7th most significant motivators, respectively. This illustrates the dependence and driving power of each category of motivators.

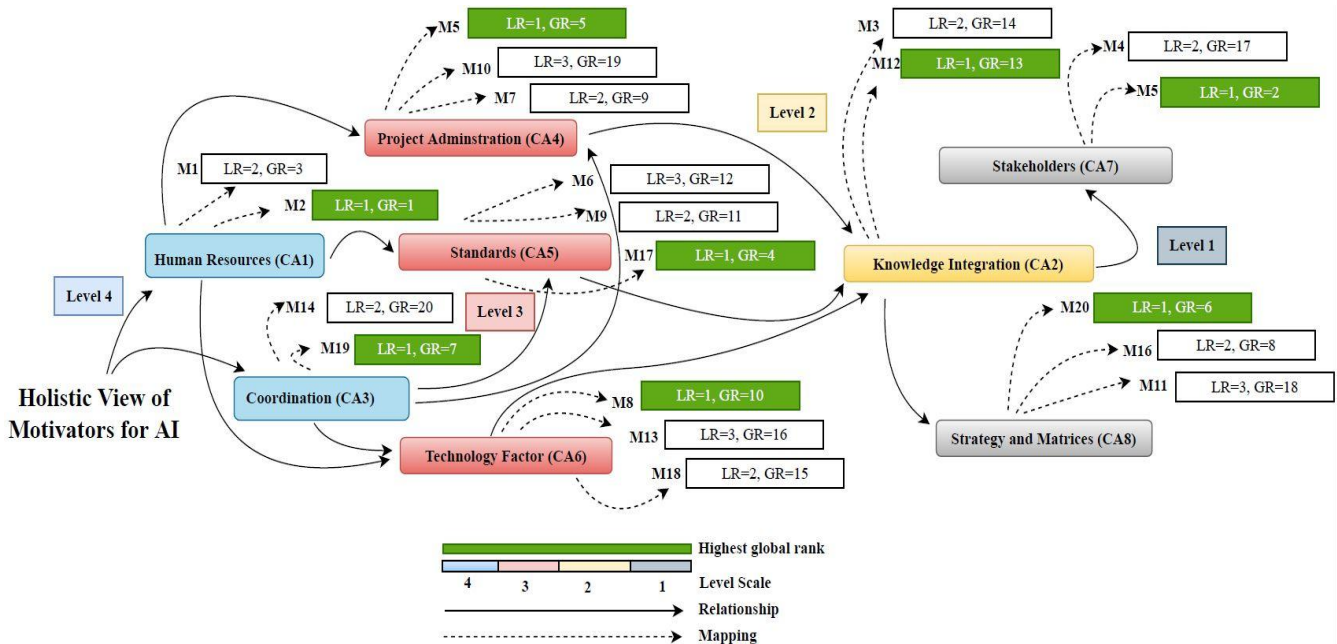

Fig 4: Holistic model of AI Ethics motivators

The holistic model showed that the stakeholder's category (CA7) is fully dependent (level 1), and the motivator "Set up a governance body to implement measures of accountability" (M15) is ranked as the 2nd highest priority motivator for considering ethics in AI-based system development. In addition, the standards category (CA5) is at level 3, and the motivator "Controllable AI with clear accountability" (M17) is ranked as the 4th most significant motivator. The motivator "Surveillance practices for data gathering and privacy of court users" (M5) is ranked 5th, and its core category, project administration (CA4), is at level 3. Hence, the holistic model provides a comprehensive framework for considering ethics in AI-based system development, with a prioritized list of motivators based on their significance. It can guide developers and stakeholders in creating ethically robust AI systems.

## 5. Conclusions, Implications, Future Of This Study And Threats To Validity

### A. Conclusions

The aim of the study was to explore and analyze the motivators of ethical considerations in AI-based system development, as reported by academic and industry researchers, as well as faced by real-world practitioners. To realize it, a multivocal literature review was conducted to identify the list of 20 motivators related to eight key categories of AI ethics. The review process involved collecting potential literature published as scientific research at academic outlets, as well as grey literature, including experience reports, blogs, white papers, case studies, etc. A total of 46 academic and 41 grey literature pieces were selected for the final data extraction process, resulting in a comprehensive list of motivators regarding industry practices. Next, a questionnaire survey was conducted with industry experts to evaluate the significance of the identified motivators in business practice. 113 complete responses got collected with the survey, and frequency analysis showed that the identified list of 20 motivators and their core categories were relevant to industry practices. The ISM approach examined the relationships between the ten core categories of AI ethics motivators. The results showed that the CA1 (Human resource) and CA3 (Coordination) categories ranked as the top (level 4), indicating that these core categories have zero dependence but strong driving power for the rest of the AI ethics motivators categories. CA4 (Project administration), CA5 (Standards), and CA6 (Technology factor) were at level 3, which suggests that these categories have both dependence (level 4) and driving relationships (for level 2 and 1). Additionally, CA1 (Human resource) and CA3 (Coordination) were fully independent categories of AI ethics motivators, whereas Stakeholders (CA7) and Strategy and matrices (CA8) were fully dependent on all other AI ethics categories.

The fuzzy TOPSIS approach was used to analyses the multicriteria decision-making problems of the identified 20 motivators by ranking them based on their significance for considering ethics in AI-based software development. The results showed that M2 (Build diverse teams), M15 (Set up a governance body to implement measures of accountability), M1 (Assign a leader responsible and accountable for ethical AI), M17 (Controllable AI with clear accountability), and M5 (Surveillance practices for data gathering and privacy of court users) were the top-ranked motivators for considering ethics in AI-based system development. To conclude, our study provides a comprehensive framework for understanding the motivators of ethics consideration in AI-based systems, with a list of prioritized motivators based on their significance. Practitioners and researchers can use it to guide the development of ethically robust AI systems.

### B. Implications

**For researchers**

This study provides a thorough review of current literature related to AI ethics, and identifies motivators that can improve the consideration of ethics in the development of AI-based systems. Our findings contribute to the existing knowledge in this subject and can assist researchers in developing strategies for creating ethically robust AI systems. Additionally, our work provides a framework for ranking the identified motivators based on their priority, and explores the relationships between the core categories of these motivators. We recommend that researchers use this prioritization-based ranking approach to direct the focus of future research on the most significant motivators for ethical AI development.

**For Practitioners**

Through a comprehensive multivocal literature review and empirical investigations, our study sets industry experts with knowledge about the motivators that can positively impact the consideration of ethics in AI-based systems. This work identifies 20 motivators that call for attention of practitioners in the development of AI-based systems. Prioritizing these motivators shall assist industry professionals in focusing on the most significant motivators first, enabling them to revise and develop new strategies for considering ethics at the development of AI-based systems. Furthermore, project managers can use the identified motivators to improve their capabilities in developing ethically robust AI systems. This can be achieved by providing training and workshops for practitioners where further skill development is needed. Top executives can also use the identified motivators and their core categories as an indicator in building necessary AI ethics capabilities, and as a risk mitigation strategy.

Additionally, the holistic model we developed can assist software providers in measuring their ability to consider ethics in AI-based system development. The model allows organizations to identify their current ethics-related strengths and weaknesses, and can be used to better plan and manage improvements, based on the insights from the developed holistic model of AI ethics motivators.

**Future of this study**

This research project aims to develop a readiness model for AI ethics (RM-AIE) to assist software development organizations in assessing their readiness regarding AI ethics implications in their development process. Additionally, this model will provide guidelines to improve their capabilities for developing ethically robust AI systems. There is a lack of established models and strategies for ethical consideration in AI-based systems. Hence, our study identified and analyzed motivators (success factors) for developing the proposed RM-AIE. The RM-AIE comprises three core components: the readiness level component, the factors component (critical success factors (CSFs), critical challenges (CCHs)), and the assessment component. The relation between these key components of RM-AIE is shown in Figure 5. The readiness level component is used to assess an organization or team's readiness level regarding the assessment and implementation of ethics in AI systems. The factor component consists of the CSFs and CCHs that represent the key areas of RM-AIE. Lastly, the assessment component is used to assess the specific readiness level of an organization and provide suggestions for best practices to improve the implication of ethical aspects in AI systems.

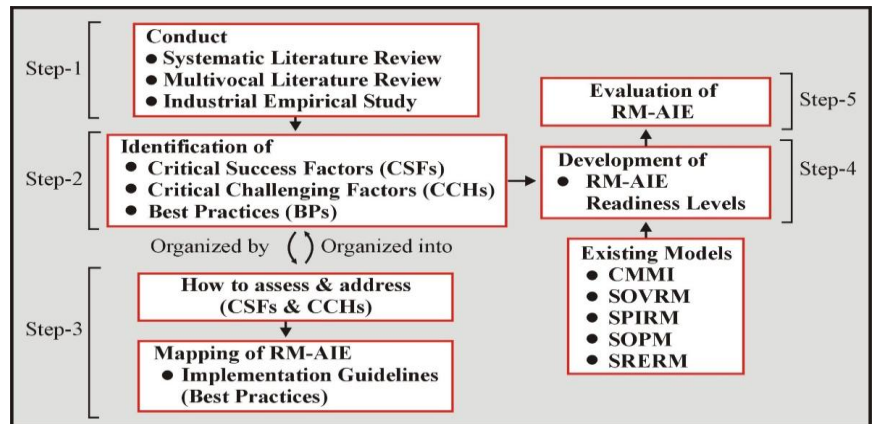

Fig 5: Structure of the proposed RM-AIE model

Moving forward, we plan to empirically explore challenging factors, best practices, and additional motivators of AI ethics. Based on our findings, we will develop the readiness levels of RM-AIE. Once the readiness levels are designed, we will conduct case studies with small, medium, and large-scale software development organizations to validate and improve the proposed model RM-AIE. We believe that the final version of RM-AIE will assist software development organizations in assessing and improving their processes for developing ethically robust AI systems. The empirical exploration of challenging factors, best practices, and motivators of AI ethics will significantly contribute to the field, which currently lacks established models and strategies for the consideration of ethics in AI-based systems.

### C.  *Validity threats and limitations*

There are potential concerns to the validity of this study. However, we took steps to alleviate these threats by following the multivocal literature review guidelines proposed by Garousi *et al.* [23], fuzzy TOPSIS developed by Hwang and Yoon [57], and ISM approach developed by Sage [51].

To collect the academic and grey literature related to the objective of our study, we performed a systematic literature review (SLR) using the multivocal literature review (MLR) method. Whereas the first author executed the search and collection process, the second author reviewed the selected literature and performed data extraction, which may have introduced bias. To address this concern, authors three and four carried out the inclusion, exclusion, quality assessment, and data extraction process for five formal and five grey studies. We also applied an inter-rater reliability test with external experts, which showed no significant bias, ensuring the consistency of our data and analysis. However, there is always a chance of missing some relevant literature during the data collection. Whereas our study has a representative set of 87 literature items, this limitation is not systematic [31, 83].

We used a questionnaire survey approach to investigate the identified motivators with industry experts. There is always a threat of bias in survey instrument development, but we addressed this by piloting and assessing the development questionnaire with external experts and colleagues. Furthermore, the findings of the Interpretive Structural Modelling (ISM) and fuzzy TOPSIS approach limited to the responses of ten experts, which may be considered a small sample. However, our study represents a subjective perspective based on the primary data we collected, and other existing works [53-55] that have also used small datasets. Therefore, the ISM and fuzzy TOPSIS approach results can be considered as generalizable.

**Appendixes**
**Appendix- A:** Selecte Literature (https://tinyurl.com/7smz27vw)
**Appendix- B:** Questionner Survey (https://forms.gle/WEC9prPYfPHHos3U8)
**Appendix- C:** Sample fo fuzzy TOPSIS Questionner( https://tinyurl.com/bdzztnnx)

**References**
[1]     M. Kuziemski and G. J. T. p. Misuraca, "AI governance in the public sector: Three tales from the frontiers of automated decision-making in democratic settings," vol. 44, no. 6, p. 101976, 2020.
[2]     V. C. Müller, "Ethics of artificial intelligence and robotics," 2020.
[3]     A. A. Khan *et al.*, "AI Ethics: An Empirical Study on the Views of Practitioners and Lawmakers," *IEEE Transactions on Computational Social Systems,* 2023.
[4]     V. Vakkuri, K.-K. Kemell, and P. Abrahamsson, "Implementing ethics in AI: initial results of an industrial multiple case study," in *International Conference on Product-Focused Software Process Improvement*, 2019: Springer, pp. 331-338.
[5]     D. Greene, A. L. Hoffmann, and L. Stark, "Better, nicer, clearer, fairer: A critical assessment of the movement for ethical artificial intelligence and machine learning," 2019.
[6]     J. Leikas, R. Koivisto, N. J. J. o. O. I. T. Gotcheva, Market,, and Complexity, "Ethical framework for designing autonomous intelligent systems," vol. 5, no. 1, p. 18, 2019.
[7]     T. J. M. Hagendorff and Machines, "The ethics of AI ethics: An evaluation of guidelines," vol. 30, no. 1, pp. 99-120, 2020.
[8]     S. Umbrello and I. Van de Poel, "Mapping value sensitive design onto AI for social good principles," *AI and Ethics,* vol. 1, no. 3, pp. 283-296, 2021.
[9]     A. L. Hunkenschroer and C. Luetge, "Ethics of AI-enabled recruiting and selection: A review and research agenda," *Journal of Business Ethics,* vol. 178, no. 4, pp. 977-1007, 2022.
[10]    B. Rakova, J. Yang, H. Cramer, and R. Chowdhury, "Where responsible AI meets reality: Practitioner perspectives on enablers for shifting organizational practices," *Proceedings of the ACM on Human-Computer Interaction,* vol. 5, no. CSCW1, pp. 1-23, 2021.
[11]    E. Halme, M. Jantunen, V. Vakkuri, K.-K. Kemell, and P. Abrahamsson, "Making ethics practical: User stories as a way of implementing ethical consideration in Software Engineering," *Information and Software Technology,* vol. 167, p. 107379, 2024.
[12]    A. Jobin, M. Ienca, and E. J. N. M. I. Vayena, "The global landscape of AI ethics guidelines," vol. 1, no. 9, pp. 389-399, 2019.
[13]    F. Rossi and N. Mattei, "Building ethically bounded AI," in *Proceedings of the AAAI Conference on Artificial Intelligence*, 2019, vol. 33, no. 01, pp. 9785-9789.
[14]    Z. Zhongming and L. Wei, "Beijing Academy of Artificial Intelligence," 2020.
[15]    R. Chatila, K. Firth-Butterfield, and J. C. Havens, "Ethically Aligned Design: A Vision for Prioritizing Human Well-being with Autonomous and Intelligent Systems Version 2," UNIVERSITY OF SOUTHERN CALIFORNIA LOS ANGELES, 2018.
[16]    A. Jobin, "others.: The global landscape of AI ethics guidelines," *Nature Machine Intelligence,* vol. 1, no. 9, pp. 389-399, 2019.
[17]    f. h. p. s. e. a. w. e. Vincent C. Müller. 2020. Ethics of Artificial Intelligence and Robotics. The Stanford Encyclopedia of Philosophy. Retrieved January 15.
[18]    W. B. Pekka Ala-Pietilä, Urs Bergmann, Mária Bieliková, Cecilia BonefeldDahl, Yann Bonnet, Loubna Bouarfa et al. (2018). The European Commission's highlevel expert group on artificial intelligence: Ethics guidelines for trustworthy AI. Working Document for stakeholders' consultation. Retrieved January 17, 2021 from https://ec.europa.eu/digital-single-market/en/news/ethics-guidelines-trustworthy-ai.

[19]  D. C. Alan Bundy. 2016. Preparing for the future of artificial intelligence. Executive Office of the President National Science and Technology Council Committee on Technology Washington, USA. Retrieved January 23, 2021 from https://cra.org/ccc/wpcontent/uploads/sites/2/2016/11/NSTC_preparing_for_the_future_of_ai.pdf.

[20]  J. S. Andrew McNamara, and Emerson Murphy-Hill. 2018. Does ACM's code of ethics change ethical decision making in software development? In Proceedings of the 26th ACM Joint Meeting on European Software Engineering Conference and Symposium on the Foundations of Software Engineering (ESEC/FSE 2018). Association for Computing Machinery, New York, NY, USA, 729–733. DOI:https://doi.org/10.1145/3236024.3264833.

[21]  V. Vakkuri, K.-K. Kemell, and P. Abrahamsson, "Implementing ethics in AI: initial results of an industrial multiple case study," in *Product-Focused Software Process Improvement: 20th International Conference, PROFES 2019, Barcelona, Spain, November 27–29, 2019, Proceedings 20*, 2019: Springer, pp. 331-338.

[22]  V. Vakkuri, K.-K. Kemell, and P. Abrahamsson, "AI ethics in industry: a research framework," *arXiv preprint arXiv:1910.12695*, 2019.

[23]  V. Garousi, M. Felderer, and M. V. Mäntylä, "Guidelines for including grey literature and conducting multivocal literature reviews in software engineering," *Information and Software Technology,* vol. 106, pp. 101-121, 2019/02/01/ 2019, doi: https://doi.org/10.1016/j.infsof.2018.09.006.

[24]  V. Garousi, M. Felderer, and T. Hacaloğlu, "Software test maturity assessment and test process improvement: A multivocal literature review," *Information and Software Technology,* vol. 85, pp. 16-42, 2017.

[25]  R. J. Adams, P. Smart, and A. S. Huff, "Shades of grey: guidelines for working with the grey literature in systematic reviews for management and organizational studies," *International Journal of Management Reviews,* vol. 19, no. 4, pp. 432-454, 2017.

[26]  H. Zhang, M. A. Babar, and P. Tell, "Identifying relevant studies in software engineering," *Information and Software Technology,* vol. 53, no. 6, pp. 625-637, 2011.

[27]  S. Jalali and C. Wohlin, "Systematic literature studies: database searches vs. backward snowballing," in *Proceedings of the 2012 ACM-IEEE International Symposium on Empirical Software Engineering and Measurement*, 2012: IEEE, pp. 29-38.

[28]  D. Badampudi, C. Wohlin, and K. Petersen, "Experiences from using snowballing and database searches in systematic literature studies," in *Proceedings of the 19th International Conference on Evaluation and Assessment in Software Engineering*, 2015: ACM, p. 17.

[29]  C. Wohlin, "Guidelines for snowballing in systematic literature studies and a replication in software engineering," in *Proceedings of the 18th international conference on evaluation and assessment in software engineering*, 2014: Citeseer, p. 38.

[30]  W. Afzal, R. Torkar, and R. Feldt, "A systematic review of search-based testing for non-functional system properties," *Information and Software Technology,* vol. 51, no. 6, pp. 957-976, 2009.

[31]  M. Niazi, S. Mahmood, M. Alshayeb, A. M. Qureshi, K. Faisal, and N. Cerpa, "Toward successful project management in global software development," *International Journal of Project Management,* vol. 34, no. 8, pp. 1553-1567, 2016.

[32]  A. A. Khan, J. Keung, M. Niazi, S. Hussain, and A. Ahmad, "Systematic literature review and empirical investigation of barriers to process improvement in global software development: Client–vendor perspective," *Information and Software Technology,* vol. 87, pp. 180-205, 2017.

[33]  S. U. Khan, M. Niazi, and R. Ahmad, "Factors influencing clients in the selection of offshore software outsourcing vendors: An exploratory study using a systematic literature review," *Journal of systems and software,* vol. 84, no. 4, pp. 686-699, 2011.

[34]  M. Niazi *et al.*, "Challenges of project management in global software development: A client-vendor analysis," *Information and Software Technology,* vol. 80, pp. 1-19, 2016.

[35]    A. A. Khan, J. Keung, S. Hussain, M. Niazi, and S. Kieffer, "Systematic literature study for dimensional classification of success factors affecting process improvement in global software development: client–vendor perspective," *IET Software,* vol. 12, no. 4, pp. 333-344, 2018.

[36]    M. Shameem, C. Kumar, B. Chandra, and A. A. Khan, "Systematic Review of Success Factors for Scaling Agile Methods in Global Software Development Environment: A Client-Vendor Perspective," in *2017 24th Asia-Pacific Software Engineering Conference Workshops (APSECW)*, 2017: IEEE, pp. 17-24.

[37]    U. Kelle, "The development of categories: Different approaches in grounded theory," *The Sage handbook of grounded theory,* vol. 2, pp. 191-213, 2010.

[38]    P. B. Change, "SPI MANIFESTO," 2010.

[39]    P. Runeson and M. J. E. s. e. Höst, "Guidelines for conducting and reporting case study research in software engineering," vol. 14, no. 2, pp. 131-164, 2009.

[40]    S. Rafi, M. A. Akbar, S. Mahmood, A. Alsanad, A. J. J. o. S. E. Alothaim, and Process, "Selection of DevOps best test practices: A hybrid approach using ISM and fuzzy TOPSIS analysis," vol. 34, no. 5, p. e2448, 2022.

[41]    M. A. Akbar, K. Smolander, S. Mahmood, A. J. I. Alsanad, and S. Technology, "Toward successful DevSecOps in software development organizations: A decision-making framework," vol. 147, p. 106894, 2022.

[42]    M. A. Akbar *et al.*, "A fuzzy analytical hierarchy process to prioritize the success factors of requirement change management in global software development," vol. 33, no. 2, e2292, 2021.

[43]    T. Punter, M. Ciolkowski, B. Freimut, and I. John, "Conducting on-line surveys in software engineering," in *2003 International Symposium on Empirical Software Engineering, 2003. ISESE 2003. Proceedings.*, 2003: IEEE, pp. 80-88.

[44]    J. S. Molléri, K. Petersen, E. J. I. Mendes, and S. Technology, "An empirically evaluated checklist for surveys in software engineering," vol. 119, p. 106240, 2020.

[45]    V. Garousi, A. Tarhan, D. Pfahl, A. Coşkunçay, and O. J. S. Q. J. Demirörs, "Correlation of critical success factors with success of software projects: an empirical investigation," vol. 27, no. 1, pp. 429-493, 2019.

[46]    B. Kitchenham and S. L. Pfleeger, "Principles of survey research: part 5: populations and samples," *ACM SIGSOFT Software Engineering Notes,* vol. 27, no. 5, pp. 17-20, 2002.

[47]    S. Ali and S. U. Khan, "Software outsourcing partnership model: An evaluation framework for vendor organizations," *Journal of systems and software,* vol. 117, pp. 402-425, 2016.

[48]    M. A. Akbar *et al.*, "Statistical analysis of the effects of heavyweight and lightweight methodologies on the six-pointed star model," *IEEE Access,* vol. 6, pp. 8066-8079, 2018.

[49]    I. Keshta, M. Niazi, and M. Alshayeb, "Towards implementation of requirements management specific practices (SP1. 3 and SP1. 4) for Saudi Arabian small and medium sized software development organizations," *IEEE Access,* vol. 5, pp. 24162-24183, 2017.

[50]    S. Mahmood, S. Anwer, M. Niazi, M. Alshayeb, and I. Richardson, "Key factors that influence task allocation in global software development," *Information and Software Technology,* vol. 91, pp. 102-122, 2017.

[51]    A. P. Sage, "Methodology for large-scale systems," 1977.

[52]    V. Ravi and R. Shankar, "Analysis of interactions among the barriers of reverse logistics," *Technological Forecasting and Social Change,* vol. 72, no. 8, pp. 1011-1029, 2005.

[53]    G. Kannan, S. Pokharel, and P. S. Kumar, "A hybrid approach using ISM and fuzzy TOPSIS for the selection of reverse logistics provider," *Resources, conservation and recycling,* vol. 54, no. 1, pp. 28-36, 2009.

[54]    H. Sharma and A. Gupta, "The objectives of waste management in India: a futures inquiry," *Technological Forecasting and Social Change,* vol. 48, no. 3, pp. 285-309, 1995.

[55]    A. Agarwal and P. Vrat, "Modeling attributes of human body organization using ISM and AHP," *Jindal Journal of Business Research,* vol. 6, no. 1, pp. 44-62, 2017.

[56] T. Raj and R. Attri, "Identification and modelling of barriers in the implementation of TQM," *International Journal of Productivity and Quality Management,* vol. 8, no. 2, pp. 153-179, 2011.

[57] K. Yoon and C.-L. Hwang, "Manufacturing plant location analysis by multiple attribute decision making: Part I—single-plant strategy," *International Journal of Production Research,* vol. 23, no. 2, pp. 345-359, 1985.

[58] T.-Y. Chen and C.-Y. Tsao, "The interval-valued fuzzy TOPSIS method and experimental analysis," *Fuzzy sets and systems,* vol. 159, no. 11, pp. 1410-1428, 2008.

[59] S. Rafi, W. Yu, M. A. Akbar, A. Alsanad, and A. Gumaei, "Multicriteria based decision making of DevOps data quality assessment challenges using fuzzy TOPSIS," *IEEE Access,* vol. 8, pp. 46958-46980, 2020.

[60] D. Kannan, A. B. L. de Sousa Jabbour, and C. J. C. Jabbour, "Selecting green suppliers based on GSCM practices: Using fuzzy TOPSIS applied to a Brazilian electronics company," *European Journal of Operational Research,* vol. 233, no. 2, pp. 432-447, 2014.

[61] R. A. Krohling and V. C. Campanharo, "Fuzzy TOPSIS for group decision making: A case study for accidents with oil spill in the sea," *Expert Systems with applications,* vol. 38, no. 4, pp. 4190-4197, 2011.

[62] F. T. Bozbura, A. Beskese, and C. Kahraman, "Prioritization of human capital measurement indicators using fuzzy AHP," *Expert systems with applications,* vol. 32, no. 4, pp. 1100-1112, 2007.

[63] A. A. L. About Oliver Beels, "Key challenges of artificial intelligence: AI ethics and governance," 2021. [Online]. Available: https://www.businessgoing.digital/key-challenges-of-artificial-intelligence-ai-ethics-and-governance/.

[64] S. Yang, T. Li, and E. van Heck, "Information transparency in prediction markets," *Decision Support Systems,* vol. 78, pp. 67-79, 2015.

[65] J. M. Carroll and M. B. Rosson, "Participatory design in community informatics," *Design studies,* vol. 28, no. 3, pp. 243-261, 2007.

[66] "Implications Of Artificial Intelligence On Patent Law," October 2020 2020. [Online]. Available: https://thelawreporter.in/2020/10/07/implications-of-artificial-intelligence-on-patent-law/#:~:text=AI%20collides%20with%20each%20of%20the%20requirement%20of,which%20can%20be%20done%20by%20the%20human%20mind.

[67] M. Soni, "End to end automation on cloud with build pipeline: the case for DevOps in insurance industry, continuous integration, continuous testing, and continuous delivery," in *2015 IEEE International Conference on Cloud Computing in Emerging Markets (CCEM)*, 2015: IEEE, pp. 85-89.

[68] R. Attri, S. Grover, N. Dev, and D. Kumar, "Analysis of barriers of total productive maintenance (TPM)," *International Journal of System Assurance Engineering and Management,* vol. 4, no. 4, pp. 365-377, 2013.

[69] J. N. Warfield, "Developing interconnection matrices in structural modeling," *IEEE Transactions on Systems, Man, and Cybernetics,* no. 1, pp. 81-87, 1974.

[70] R. Fadilah and H. Kuswoyo, "Transitivity analysis of presidential debate between Trump and Biden in 2020," *Linguistics and Literature Journal,* vol. 2, no. 2, pp. 98-107, 2021.

[71] F. R. L. Junior, L. Osiro, and L. C. R. Carpinetti, "A comparison between Fuzzy AHP and Fuzzy TOPSIS methods to supplier selection," *Applied Soft Computing,* vol. 21, pp. 194-209, 2014.

[72] C.-N. Liao and H.-P. Kao, "An integrated fuzzy TOPSIS and MCGP approach to supplier selection in supply chain management," *Expert Systems with Applications,* vol. 38, no. 9, pp. 10803-10811, 2011.

[73] A. Zouggari and L. Benyoucef, "Simulation based fuzzy TOPSIS approach for group multi-criteria supplier selection problem," *Engineering Applications of Artificial Intelligence,* vol. 25, no. 3, pp. 507-519, 2012.

[74] D. J. T. I. J. o. A. M. T. Yong, "Plant location selection based on fuzzy TOPSIS," vol. 28, no. 7, pp. 839-844, 2006.

[75] L. Hofeditz, M. Mirbabaie, A. Luther, R. Mauth, and I. Rentemeister, "Ethics Guidelines for Using AI-based Algorithms in Recruiting: Learnings from a Systematic Literature Review," in *HICSS*, 2022, pp. 1-10.

[76] J. C. Sipior, "Considerations for development and use of AI in response to COVID-19," *International Journal of Information Management,* vol. 55, p. 102170, 2020.

[77] J. Ochmann and S. Laumer, "Fairness as a determinant of AI adoption in recruiting: An interview-based study," 2019.

[78] V. Vakkuri and K.-K. Kemell, "Implementing AI ethics in practice: An empirical evaluation of the RESOLVEDD strategy," in *Software Business: 10th International Conference, ICSOB 2019, Jyväskylä, Finland, November 18–20, 2019, Proceedings 10*, 2019: Springer, pp. 260-275.

[79] W. H. Organization, "Ethics and governance of artificial intelligence for health: WHO guidance," 2021.

[80] A. Gardner, A. L. Smith, A. Steventon, E. Coughlan, and M. Oldfield, "Ethical funding for trustworthy AI: proposals to address the responsibilities of funders to ensure that projects adhere to trustworthy AI practice," *AI and Ethics,* pp. 1-15, 2022.

[81] B. C. Stahl, J. Antoniou, M. Ryan, K. Macnish, and T. Jiya, "Organisational responses to the ethical issues of artificial intelligence," *AI & SOCIETY,* vol. 37, no. 1, pp. 23-37, 2022.

[82] M. Janssen, P. Brous, E. Estevez, L. S. Barbosa, and T. Janowski, "Data governance: Organizing data for trustworthy Artificial Intelligence," *Government Information Quarterly,* vol. 37, no. 3, p. 101493, 2020.

[83] M. A. Akbar *et al.*, "Success factors influencing requirements change management process in global software development," *Journal of Computer Languages,* vol. 51, pp. 112-130, 2019.